\documentclass{article}
\usepackage{amsmath}
\usepackage{amssymb}
\usepackage{amsthm}
\usepackage{graphicx} 
\usepackage{natbib}
\usepackage{enumitem}
\usepackage{wrapfig}
\usepackage{caption}
\usepackage{booktabs}
\usepackage{tikz}

\newcommand\independent{\protect\mathpalette{\protect\independenT}{\perp}}
\def\independenT#1#2{\mathrel{\rlap{$#1#2$}\mkern2mu{#1#2}}}

\newcommand{\N}{\mathbb{N}}
\newcommand{\lep}{\mathrel{\overset{+}{\le}}}   
\newcommand{\eqp}{\mathrel{\overset{+}{=}}}    
\newcommand{\gep}{\mathrel{\overset{+}{\ge}}}   
\newcommand{\pair}[1]{\langle #1 \rangle}

\theoremstyle{plain}
\newtheorem{theorem}{Theorem}
\newtheorem{proposition}{Proposition}
\newtheorem{lemma}{Lemma}
\newtheorem{corollary}{Corollary}
\newtheorem{question}{Question}
\newtheorem{conjecture}{Conjecture}
\theoremstyle{definition}
\newtheorem{definition}{Definition}
\theoremstyle{remark}
\newtheorem{remark}{Remark}

\title{Copying versus self-construction in cellular automata: an algorithmic-information approach to self-reproduction}
\author{Konstantin Doubrovinski \thanks{correspondence: Konstantin.Doubrovinski@UTSouthwestern.edu} }
\date{\today}

\begin{document}

\maketitle

\begin{abstract}
What distinguishes an object that reproduces itself from one that is merely copied by its environment? We propose an answer based on algorithmic information and interventions, in cellular automata. Copies of an object are identified by the algorithmic information they share with it. Its code and its machine are identified by perturbations: changes to the code are inherited by the offspring, changes to the machine are not, since the offspring’s machine is rebuilt from the code. An object is a self-reproducer if it proliferates, admits such a decomposition, its machine can be decoded from its code, and removing the machine stops the code from being copied. The definition uses only counts, interventions and upper bounds on information, so it is not undermined by exhaustive search. We prove that it excludes trivial “infection” dynamics and rules that copy every pattern, and reduce the claim that it includes the construction of von Neumann and Burks, but not its constituents, to explicit hypotheses. We measure the amount of self-reproduction by the length of the shortest proof, in a fixed formal theory, that some machinery can stand in for the object’s machine; its lower bounds are decidable and immune to search. In an abstract model of reproduction we show that both ingredients are necessary: no criterion computed from the observed run alone distinguishes a machine built from its description from one that is copied, and criteria that ignore the dynamical law attribute to the object information the law supplies. We also show that conditional Kolmogorov complexity is, up to logarithmic terms, the only effective notion of determination compatible with counting and with charging re-encodings by their description length. Finally, we show that separate weak dependence of a function on its arguments implies joint weak dependence only non-effectively, with a loss governed by the halting problem.
\end{abstract}

\section{Introduction}

Self-reproduction is easy to recognize in examples and difficult to define. A definition should include the self-reproducing machine of von Neumann and Burks, and it should exclude processes in which an object is merely copied by its environment, or in which ``reproduction'' amounts to the spreading of a trivial pattern. It should also be insensitive to the details of implementation. The purpose of this work is to propose such a definition, using algorithmic information theory to express causal relations between individual objects.

We begin by introducing the necessary notation and reviewing several notions from algorithmic information theory, Judea Pearl causality, and some mathematical constructions formalizing the notion of self-reproduction (autopoiesis).

\textit{Algorithmic Information Theory and Kolmogorov Complexity}: We assume familiarity with basic notions from classical Recursion Theory and Algorithmic Information Theory (for excellent sources on this subject see \cite{li2008} and references therein). Here, a string will refer to a finite binary sequence. Fixing some computable bijection between strings and natural numbers, e.g. lexicographical ordering, we can refer to them interchangeably. It is also useful to introduce a pairing function, i.e. a recursive bijection between strings and their ordered pairs. In this way, every ordered pair of strings $x$, $y$ will correspond to a unique string $\langle x,y \rangle$, and vice versa. (Cantor pairing function is a popular choice, sending a pair of numbers $\langle x, y\rangle$ to $(x^2+x+2xy+3y+y^2)/2$). Having fixed a pairing function, there is no need to distinguish between computable functions of one argument and computable functions of two arguments (and this can be extended to tuples in the obvious way). Let us fix a universal Turing machine, or ``mode of description'', $U$, to define a G\"odel numbering of partial recursive functions. More specifically, for every partial recursive function of two arguments $V(p,x)$, there exists a total recursive $s$, such that $V(p,x)=U(s(p),x)$. We can refer to $s$ as an ``interpreter'' translating between ``programming languages'' $V$ and $U$. Throughout, we take $U$ to be \emph{prefix-free}, i.e. the set of programs $p$ on which $U(p,y)$ halts is, for every fixed $y$, prefix-free (no halting program is a proper prefix of another), and universal among such machines. The Kolmogorov complexity, or algorithmic entropy, of a string is the length of the shortest program that outputs it:
\[
K(x) \equiv \min \{ l(p) \, \lvert \,  U(p)=x \} \quad,
\]
where $l(p)$ is the length of the string $p$, and $U(p)$ abbreviates $U(p,\varepsilon)$ with $\varepsilon$ the empty string. Similarly, the conditional complexity of $x$ given $y$ is $K(x \, | \, y) \equiv \min \{ l(p) \, | \, U(p,y)=x\}$. (With a prefix-free $U$, $K$ is usually called the prefix complexity; it differs from the complexity defined with an arbitrary universal machine by at most $O(\log l(x))$, and is technically more convenient below.) We denote by $x^*$ the first shortest program for $x$ (in the order in which programs of length $K(x)$ are found to halt); $x^*$ and $\langle x, K(x) \rangle$ are computable from each other. Kolmogorov complexity can be assigned to any finitely describable object by putting the set of those objects in one-to-one correspondence with the natural numbers. In what follows, we will also be considering Kolmogorov complexity of recursive sets that may or may not be finite. Kolmogorov complexity of a recursive set $S$ will be the complexity of the (simplest G\"odel number of) the recursive function computing its characteristic sequence:
\begin{align*}
K(S) &\equiv \min \{ K(i) \, | \, \varphi_i \text{ is total and } \varphi_i = \chi_S \} \quad, \\
K(S \, | \, y) &\equiv \min \{ K(i \, | \, y) \, | \, \varphi_i \text{ is total and } \varphi_i = \chi_S \} \quad,
\end{align*}
where $\varphi_i$ is the $i$-th partial recursive function and $\chi_S$ is the characteristic function of $S$.

Recursive isomorphism is a total recursive bijection from natural numbers to natural numbers. A property (of a set of natural numbers) is said to be recursively invariant if it is preserved by recursive isomorphisms. For example, being finite or being recursive are recursively invariant notions. Viewing a string as a singleton set, its Kolmogorov complexity is not strictly recursively invariant. However, it is approximately invariant in the following strong sense: when changing the mode of description from $U_1$ to $U_2$, the Kolmogorov complexity of $x$ changes by no more than a constant depending on $U_{1,2}$, but not on $x$. This is readily derived from the fact that any program in one Turing-complete language can be transformed into a corresponding program in another language, thus only requiring an additional fixed overhead to encode the interpreter.

\textit{Notation for approximate (in)equalities and standard facts}: For quantities $a$ and $b$ depending on strings, we write $a \lep b$ if there is a constant $c$ such that $a \le b + c$ for all values of the strings involved; $a \eqp b$ means $a \lep b$ and $b \lep a$. The constant $c$ may depend on $U$ and on the computable functions explicitly fixed in the statement (such as $f$, $g$, $h$ below), but never on the strings. We shall use the following standard facts (see \cite{li2008, gacs1974, chaitin1975}). For all strings $x$, $y$, $z$:
\begin{enumerate}[label=(F\arabic*)]
\item $K(x) \lep l(x) + 2 \log l(x)$ and $K(x \, | \, y) \lep K(x)$.
\item For every computable $g$, $K(g(x,y) \, | \, y) \lep K(x \, | \, y)$ (run a program for $x$ given $y$, then apply $g$). In particular, $K(\langle x,y \rangle \, | \, z) \lep K(x \, | \, z) + K(y \, | \, z)$ (concatenate two self-delimiting programs), and $K(x \, | \, \langle y, z\rangle) \lep K(x \, | \, y)$.
\item (Symmetry of information, exact form \cite{gacs1974, chaitin1975}.) $K(\langle x,y \rangle) \eqp K(x) + K(y \, | \, x^*)$. Moreover $K(y \, | \, x^*) \lep K(y \, | \, x)$, since $x$ is computable from $x^*$.
\item (Symmetry of information, conditional form.) There is a constant $c$ such that $\bigl| K(\langle x,y \rangle \, | \, z) - K(x \, | \, z) - K(y \, | \, \langle x,z \rangle) \bigr| \le c \log (n+2)$, where $n = l(x)+l(y)+l(z)$.
\end{enumerate}
Following \cite{li2008, janzing2010}, the (algorithmic) mutual information and the conditional mutual information are
\begin{gather*}
I(x:y) \equiv K(x) + K(y) - K(\langle x,y \rangle) \quad , \\ I(x:y \, | \, z) \equiv K(x \, | \, z) + K(y \, | \, z) - K(\langle x,y \rangle \, | \, z) \quad .
\end{gather*}
By (F2), $I(x:y) \ge -c$ and $I(x:y\,|\,z) \ge -c$ for a constant $c$, and both are symmetric up to $O(1)$, since $K(\langle x,y\rangle \,|\, z) \eqp K(\langle y,x\rangle \,|\, z)$.

\textit{Judea Pearl Causality}: To introduce the notion of causality, consider a set of stochastic variables $X$, $Y$, $Z$,... with a joint probability distribution $P(X,Y,Z,...)$. We say that $X$ is independent of $Y$ given $Z$, written as $X \independent Y \, \lvert \, Z$, to mean $P(X,Y,Z) = h(X,Z)k(Y,Z)$, for some functions $h$ and $k$. In other words, informally, if given $Z$, $Y$ is of no use to predict the value of $X$. Thus, conditional independence is a relation on triplets of stochastic variables, which may be shown to have the following properties \cite{dawid1979, pearlpaz1987, pearl2009}:
\begin{center}
\small
$\begin{array}{l}
(C1) \quad  X \independent Y \, \lvert \, Z \implies Y \independent X \, \lvert \, Z \\
(C2) \quad  X \independent Y \, \lvert \, Z \; \text{and} \;  U=h(X) \implies U \independent Y \, \lvert \, Z \\
(C3) \quad  X \independent Y \, \lvert \, Z \; \text{and} \;  U=h(Y) \implies X \independent Y \, \lvert \, (Z,U) \\
(C4) \quad  X \independent Y \, \lvert \, Z \; \text{and} \; X \independent W \, \lvert \, (Y,Z) \implies X \independent (W,Y) \, \lvert \, Z \\
(C5) \quad  X \independent Y \, \lvert \, Z \; \text{and} \; X \independent Z \, \lvert \, Y \implies X \independent (Y,Z) \quad,
\end{array}$
\end{center}
\noindent where $(C5)$ holds when there is no non-trivial relationship between $Y$ and $Z$ (for instance, when $P$ is strictly positive; the precise condition is given in Proposition \ref{prop:prob-intersection} in Appendix \ref{app:further}). Properties (C1)--(C4) are known as the semi-graphoid axioms, and (C1)--(C5) as the graphoid axioms \cite{pearlpaz1987}. In Pearl's approach, causal structure is reflected in, and to a large extent inferred from, such conditional independences. Somewhat informally, $A$ is the cause of $B$ if $B$ is independent of all else, provided $A$ is given. In addition, causal claims can be tested by interventions, i.e. by setting a variable to a prescribed value from outside the system and observing the effect (the ``do-operator'' \cite{pearl2009}). Remarkably, the same axioms apply to graph separation in the following sense. Given a graph, consider three subsets of its vertices $X$, $Y$, and $Z$. Consider the following relation: any path from $X$ to $Y$ goes through $Z$. This relation conforms to axioms (C1)-(C5), provided we modify (C2) and (C3) to mean $U \subseteq X$ (respectively $U \subseteq Y$), and only stipulate (C5) under the assumption that $X$, $Y$, and $Z$ are disjoint. Axioms (C1)-(C5) apply in contexts unrelated to graphs and probability distributions. An abstract setting for them is Dawid's notion of a separoid \cite{dawid2001}: a ternary relation on a join semilattice $(V, \le)$, where ``$U = h(X)$'' is replaced by ``$U \le X$'' and the pair $(Y,Z)$ by the join $Y \vee Z$; see also \cite{studeny2005}. Janzing and Sch\"olkopf \cite{janzing2010} proposed replacing stochastic conditional independence by the vanishing of algorithmic conditional mutual information. This makes it possible to speak of causal relations between individual objects, such as two strings or two patterns of cells, rather than between random variables. The present work builds on this idea.

\textit{Self-replicating machines}: Finally, let us briefly recall some work on formalizing the notion of self-reproduction. In von Neumann's influential book ``Theory of Self-Reproducing Automata'', completed and published in 1966 by Arthur W. Burks, the authors present a specific construction meant to represent a genuinely self-reproducing machine modeled within a cellular automaton environment. A cellular automaton may be informally described as an infinite grid of cells (two-dimensional in the work of von Neumann and Burks), where each cell can assume some finite number of states (29 in the said work). Time-dynamics are discrete with each cell changing state depending on its current state and the states of its adjacent neighboring cells. The von Neumann--Burks construction is rather nuanced, but the key features may be summarized as follows (after Odifreddi, ref ??). The cellular environment can harbor a set of ``machines'' represented by (connected) regions of cellular space in which cells are initiated in some specific states. Among these machines there is one with the following property: given input ``instructions'' describing another machine, it ``builds'' that machine according to the provided instructions. This machine is called the universal constructor, and its functioning can be summarized as $(A,d_X) \rightarrow X$, to signify that machine $A$ builds machine $X$ given instructions $d_X$. All machines harbored by the environment are assumed to have some description that may be provided to the universal constructor $A$. (This in essence implies Turing completeness: the environment can emulate the universal Turing machine defined in the beginning of the Introduction section). Additionally, the environment harbors machine $B$ that produces a copy of any instructions when given those instructions as input, i.e. $(B,d_X) \rightarrow d_X$. Finally, machines $A$ and $B$ can be combined into a single machine $D$ that on input $d_X$ first schedules machine $A$ to produce machine $X$, next $D$ schedules machine $B$ to also replicate a copy of the instructions $d_X$, and finally ``superimposes'' the two to produce machine $X$ ``pre-loaded'' with input $d_X$. Hence, $(D,d_X) \rightarrow (X,d_X)$. (In this summary, a machine that has finished its task persists, so that, for instance, $(D,d_X) \rightarrow (X,d_X)$ leaves $D$ in place next to the newly built pair.) As was said, all machines have some description, including machine $D$. Formally substituting $D$ for $X$ in the last expression, we get $(D,d_D) \rightarrow (D,d_D)$, such that machine $D$ copies both itself and its description, hence completing a round of self-replication. The next round of replication is then initiated by the newly replicated copy of machine $D$. In this way, the number of machines in the cellular environment grows without bound with increasing time. Note that the constituents of $D$ also produce copies of something: $(A, d_A) \rightarrow A$, and $(B, d_X) \rightarrow d_X$ for every $X$. Neither is regarded as self-reproduction. The constructor $A$ builds a copy of itself only if it is given its own description, which it does not copy, so its offspring cannot reproduce further without a new description supplied from outside. The copier $B$ copies any description given to it, so the copying is performed by $B$ and not by the description being copied. The von Neumann--Burks construction is rather involved. Subsequently, considerably simpler constructions of ``self-replicating machines'' were proposed, notably Langton's self-replicating loops. These do not (obviously) require Turing completeness of the cellular environment and do not (necessarily) harbor a universal constructor. Regressing to absurdity, one may consider a 2D environment where cells can have one of two states, $0$ and $1$, with an update rule switching $0$ to $1$ if one of a cell's neighbors is in state $1$. (In this way, a cell ``reproduces" by ``infecting" its neighbors). This automaton will thus have $(t+1)^2$ cells in state $1$ after $t$ time steps. Thus, a definition of self-reproduction should include $D$, and exclude its constituents $A$ and $B$, the ``infection'' automaton, and rules of dynamics that copy every pattern (an example is given in Section \ref{sec:copies}).

It is also interesting to mention an alternative approach to self-reproduction that has recently emerged in the Soft-Matter Physics community. There researchers often aim to develop an actual physical self-reproducing entity, using e.g. DNA origami, rather than a mathematical model thereof, ref ??. It was proposed that self-reproduction should be defined as a property of the system that after time $t$ produces exponentially many ($ O(2^ {\lambda t})$, $\lambda > 0$) copies of some object if ``building blocks'' comprising those objects are provided in excess.

In view of the last part of our introductory discussion it is clear that the current notion of self-replication is highly informal and one may even wonder if such a definition is of any ``serious'' mathematical interest. The reason for pessimistic (dismissive?) attitudes may come from clear lack of invariance in the above definitions. When definitions rely on very particular constructions (such as 2D cellular automata), they inevitably harbor a good deal of ambiguity. Here, we wish to overcome this issue. We will construct an Algorithmic-Information-Theory-based definition of self-replication which (to an extent) overcomes ambiguities of implementation by instead focusing on how information within the self-replicating system is being exchanged.

\textit{Outline and contributions}: In Section 2 we introduce a notion of weak dependence of a computable function on its argument. We show that it coincides with the algorithmic mutual information between argument and value (Proposition \ref{prop:collapse}), so that it is an instance of algorithmic conditional independence in the sense of \cite{janzing2010}. For functions of two arguments, we show that weak dependence on each argument separately implies weak dependence on both (Theorem \ref{thm:one}). The bound holds only up to the mutual information of the arguments (Theorem \ref{thm:effective}); without this term, the best bound grows faster than any computable function (Proposition \ref{prop:busybeaver}), although it becomes effective relative to the halting problem (Proposition \ref{prop:relativized}). We also give self-contained proofs that algorithmic conditional independence satisfies the graphoid axioms with explicit losses (a known fact \cite{janzing2010, steudel2010}), and, in Appendix \ref{app:further}, make precise an analogy with differentiation. We also show that conditional Kolmogorov complexity is, up to logarithmic terms, the only effective notion of ``$y$ is determined by $x$ up to $k$ bits'' that respects counting and charges computable re-encodings by their description length (Theorem \ref{thm:determination}). In Section 3 we apply these notions to cellular automata. We express causal relations between regions of cellular space as conditional mutual informations, identify copies of an object by the information they share with it (Proposition \ref{prop:tc}), and identify the code and the machine of an object by interventions, as its heritable and non-heritable parts. On this basis we propose a definition of self-reproduction (Definition \ref{def:selfrep}) and two measures of its amount, an algorithmic one (Definition \ref{def:amount}) and a certified one, the length of the shortest proof in a fixed formal theory that some machinery can do the job of the object's machine (Definition \ref{def:certified}), and prove that the definition excludes the infection automaton and rules that copy every pattern (Proposition \ref{prop:exclusions}). We present both measures as minimum description length, in an untyped and a typed system of descriptions respectively, and state the thesis that the section supports, together with the alternatives it rules out (Section \ref{sec:mdl}). We then show, in an abstract model, that both ingredients of the definition are necessary: interventions (Proposition \ref{prop:observational}) and information relative to the dynamical law (Proposition \ref{prop:keys}). We explain why the algorithmic measure is vulnerable to exhaustive search (Remark \ref{rem:search}) and why the certified measure is not (Remark \ref{rem:reflection}), and reduce the claim that the definition includes the construction of von Neumann and Burks to explicit hypotheses (Proposition \ref{prop:exact} and Conjecture \ref{conj:vonneumann}). The material of Section 2 that is not needed in Section 3 is collected in Appendix \ref{app:further}, and an alternative formulation of Section 3, based on sorted automata, in Appendix \ref{app:sorted}.


\hfill \\

\section{Algorithmic notion of weak dependence.}

\subsection{Definitions}

Consider computable function $f(x)$. $f$ is independent of its argument usually simply means that $f$ is constant. This notion is a special (trivial) case of independence in the sense of axioms (C1)-(C5). However, it is too strict for our present purposes. We thus weaken it by making it quantitative instead of qualitative, at the same time preserving recursive invariance. Specifically, fix computable $f(x)$ and some value of its argument $x=x_0$. Consider sets which map to $f(x_0)$ under $f$ and contain $x_0$ (i.e. all $S$ such that  $S \subseteq f^{-1}(f(x_0))$ and $x_0 \in S$). We will say that $f$ depends weakly on $x$ at $x_0$ if the Kolmogorov complexity of the simplest set in this class is small. In other words, for $f(x)$ to assume the value $f(x_0)$, it suffices that $x$ has some simple property, also true of $x_0$ ($x_0$ is a member of a simple set, e.g. being even or prime). Recall that we assume $S$ is recursive, but not necessarily finite, having Kolmogorov complexity $ K(S) = \min_i K(i)$ such that $\varphi_i$ computes the characteristic function of $S$. The derivative notation below is motivated in Appendix \ref{sec:derivative}. Throughout this section, ``computable function'' means \emph{total} computable function.

\begin{definition}[Weak dependence]\label{def:weak}
Let $f : \N \to \N$ be computable and $x_0 \in \N$. Set
\[
f'(x) \, |_{x_0}  \equiv \min \{ K(S) \, | \, S \text{ recursive}, \; x_0 \in S \subseteq f^{-1}(f(x_0)) \} \quad .
\]
We say that $f$ depends weakly on $x$ at $x_0$ if $f'(x)\,|_{x_0}$ is small.
\end{definition}

The minimum is over a non-empty class: since $f$ is total, the level set $f^{-1}(f(x_0))$ is itself recursive, and it contains $x_0$. We will also consider functions of several variables $f(x,y,...)$, with e.g.
\begin{equation}
\partial_{x} f(x,y) \, |_{x_0,y_0}  \equiv \min \bigl\{ K(S) \, \big| \, x_0 \in S, \; f(S \times \{ y_0 \}) = \{ f(x_0,y_0) \} \bigr\} \quad ,
\label{eq:weak}
\end{equation}
where $S$ ranges over recursive subsets of $\N$,
i.e. the preimage is taken w.r.t. $x$, while $y=y_0$ is held fixed; $\partial_{y} f(x,y) \, |_{x_0,y_0}$ is defined symmetrically. We also introduce the \emph{conditional} variant, in which the set $S$ may be described using the frozen argument:
\begin{equation}
\partial^{\,|}_{x} f(x,y) \, |_{x_0,y_0}  \equiv \min \bigl\{ K(S \, | \, y_0) \, \big| \, S \text{ as in (\ref{eq:weak})} \bigr\} \quad ,
\label{eq:weakcond}
\end{equation}
and symmetrically $\partial^{\,|}_{y} f$, with $K(S \,|\, x_0)$. Finally, $\partial_{x,y} f \, |_{x_0,y_0} < c$ is shorthand for ``$\partial_{x} f \, |_{x_0,y_0} < c$ and $\partial_{y} f \, |_{x_0,y_0} < c$''.

\begin{remark}\label{rem:sets}
In (\ref{eq:weak}) the set $S$ is a set of first coordinates, i.e. $S \subseteq \N$ with $S \times \{y_0\} \subseteq f^{-1}(f(x_0,y_0))$. If instead one minimized over sets of pairs $S' \subseteq \N \times \{y_0\}$, then always $K(S') \gep K(y_0)$ (given $S'$, find its least element and output its second coordinate), so the partial derivative would be dominated by the complexity of the frozen argument rather than by the dependence of $f$ on $x$. This is why (\ref{eq:weak}) uses $S \subseteq \N$, which is also the form used in the proof of Theorem \ref{thm:one} below.
\end{remark}

\subsection{Weak dependence measures information transmitted from argument to value}

The following simple observation identifies the quantities just defined.

\begin{proposition}\label{prop:collapse}
For every computable $f$ there is a constant $c$ such that for all $x_0$,
\[
\bigl| \, f'(x)\,|_{x_0} - K(f(x_0)) \, \bigr| \le c \quad , \qquad \text{i.e.} \quad f'(x)\,|_{x_0} \eqp K(f(x_0)) \quad.
\]
\end{proposition}
\begin{proof}
Write $v = f(x_0)$. \emph{Upper bound.} By the $s$-$m$-$n$ theorem there is a total computable $s$ such that, for every $w$, $\varphi_{s(w)}$ is the characteristic function of $f^{-1}(w)$ (on input $x$, compute $f(x)$ and compare it with $w$; this is total because $f$ is). The set $f^{-1}(v)$ is admissible in Definition \ref{def:weak}, so $f'(x)\,|_{x_0} \le K(s(v)) \lep K(v)$ by (F2).
\emph{Lower bound.} Let $S$ be admissible and let $i$ be an index with $\varphi_i = \chi_S$ and $K(i) = K(S)$. Consider the following procedure: given $i$, compute $\varphi_i(0), \varphi_i(1), \ldots$ until the first $s$ with $\varphi_i(s)=1$ is found (it exists since $x_0 \in S$, and every $\varphi_i(n)$ is defined since $\varphi_i$ is total); output $f(s)$. Since $s \in S \subseteq f^{-1}(v)$, the output is $v$. Hence $K(v) \lep K(i) = K(S)$, for every admissible $S$.
\end{proof}

\begin{corollary}\label{cor:mi}
$f'(x)\,|_{x_0} \eqp I(x_0 : f(x_0))$.
\end{corollary}
\begin{proof}
By (F2), $K(\langle x_0, f(x_0) \rangle) \eqp K(x_0)$ (each of $x_0$ and $\langle x_0, f(x_0)\rangle$ is computable from the other). Hence $I(x_0 : f(x_0)) = K(x_0) + K(f(x_0)) - K(\langle x_0,f(x_0)\rangle) \eqp K(f(x_0))$, and Proposition \ref{prop:collapse} applies.
\end{proof}

Thus the weak dependence of $f$ on $x$ at $x_0$ is the amount of algorithmic information about $x_0$ that is transmitted to the value $f(x_0)$. For example, if $f$ is constant, or $f(x) = x \bmod 2$, then $f'(x)\,|_{x_0} \eqp 0$ for all $x_0$; if $f$ is the identity, then $f'(x)\,|_{x_0} \eqp K(x_0)$. Note also that the ``simple property'' of $x_0$ in the informal description above may always be taken to be the property ``$f(x) = f(x_0)$'' itself. The same reasoning applies to partial derivatives.

\begin{proposition}\label{prop:partial}
For every computable $f(x,y)$, for all $x_0,y_0$, with $v = f(x_0,y_0)$:
\begin{enumerate}[label=(\roman*)]
\item $K(v \, | \, y_0) \lep \partial_x f \,|_{x_0,y_0} \lep K(\langle y_0, v\rangle)$;
\item $\partial^{\,|}_x f \,|_{x_0,y_0} \eqp K(v \,|\, y_0) \eqp I(x_0 : v \,|\, y_0)$.
\end{enumerate}
Symmetric statements hold for $\partial_y f$ and $\partial^{\,|}_y f$.
\end{proposition}
\begin{proof}
(i) Lower bound: given $y_0$ and an index $i$ of an admissible $S$, search for the least $s \in S$ and output $f(s,y_0) = v$; hence $K(v \,|\, y_0) \lep K(i)$. Upper bound: the set $\{ x \,|\, f(x,y_0) = v\}$ is admissible, and an index for it is computable from $\langle y_0, v\rangle$.
(ii) The lower bound $K(v\,|\,y_0) \lep \partial^{\,|}_x f$ is proved exactly as in (i), now with $K(i\,|\,y_0)$ in place of $K(i)$. For the upper bound, an index of $\{ x \,|\, f(x,y_0) = v\}$ is computable from $v$ given $y_0$, so $\partial^{\,|}_x f \lep K(v\,|\,y_0)$. Finally, $\langle x_0, v\rangle$ and $x_0$ are computable from each other given $y_0$, so $K(\langle x_0,v\rangle \,|\, y_0) \eqp K(x_0 \,|\, y_0)$ by (F2), and therefore $I(x_0 : v \,|\, y_0) \eqp K(v \,|\, y_0)$.
\end{proof}

\begin{remark}\label{rem:locality}
Because the level set itself is always admissible, Definition \ref{def:weak} does not measure whether $f$ is constant on a ``small neighbourhood'' of $x_0$; it depends on $x_0$ only through the value $f(x_0)$. A genuinely local notion would require restricting the class of admissible sets (e.g. to finite sets, with a penalty $\log |S|$, or to sets of bounded time complexity). We do not pursue this here: the identification with mutual information is precisely what allows a rigorous comparison with (C1)--(C5) in Section \ref{sec:ci}.
\end{remark}

\subsection{Composition and joint dependence}

\begin{theorem}\label{thm:one}\
\begin{enumerate}[label=(\roman*)]
\item There exists a constant $c$ such that $ \frac{d g(f(x))}{dx} |_{x_0} < \frac{d f(x)}{dx} |_{x_0} + c$, where $c$ may depend on $g$, but not on $x_0$. In fact one may take $c=1$, i.e. $\frac{d g(f(x))}{dx} |_{x_0} \le \frac{d f(x)}{dx} |_{x_0}$; moreover $\frac{d g(f(x))}{dx} |_{x_0} \eqp \frac{d g(y)}{dy}|_{f(x_0)}$.
\item For any computable $f(x,y)$ of two variables, if $\partial_{x,y} f(x,y) \, |_{x_0,y_0} < c$, then, considering $f$ a function of a single $z = \langle x, y \rangle$, $\partial_{z} f(z) \, |_{\langle x_0,y_0 \rangle} < c'$, where $c'$ may depend on $c$ and $f$, but not on $x_0$ and $y_0$.
\end{enumerate}
\end{theorem}

\noindent
Here, in (ii), ``$f$ considered as a function of $z$'' is $\tilde f(z) = f(\pi_1(z), \pi_2(z))$, where $\pi_{1,2}$ are the (computable) inverses of the pairing function, so that $\partial_z f \,|_{\langle x_0, y_0 \rangle} \equiv \tilde f'(z)\,|_{\langle x_0, y_0 \rangle}$.

The key step of the proof of (ii) is the following lemma, illustrated in Fig. \ref{fig1}. For sets $S_x, S_y \subseteq \N$ call a point $\langle x_1, y_1\rangle \in S_x \times S_y$ \emph{good} (for $f$, $S_x$, $S_y$) if $f(\{x_1\} \times S_y) = f(S_x \times \{y_1\}) = \{ f(x_1,y_1) \}$, i.e. $f$ is constant on the ``row'' and on the ``column'' of $S_x \times S_y$ passing through $\langle x_1,y_1\rangle$.

\begin{lemma}\label{lem:good}
All good points of $S_x \times S_y$ have the same value of $f$.
\end{lemma}
\begin{proof}
Let $\langle x_1,y_1 \rangle$ and $\langle x_2,y_2 \rangle$ be good. Since $y_1, y_2 \in S_y$ and $f$ is constant on $\{x_1\} \times S_y$, $f(x_1,y_1) = f(x_1,y_2)$. Since $x_1, x_2 \in S_x$ and $f$ is constant on $S_x \times \{ y_2\}$, $f(x_1,y_2) = f(x_2,y_2)$. Hence $f(x_1,y_1) = f(x_2,y_2)$: perpendicular lines through good points cross.
\end{proof}

\begin{center}
\IfFileExists{./perpendicular.pdf}{%
\includegraphics[width=0.48\textwidth]{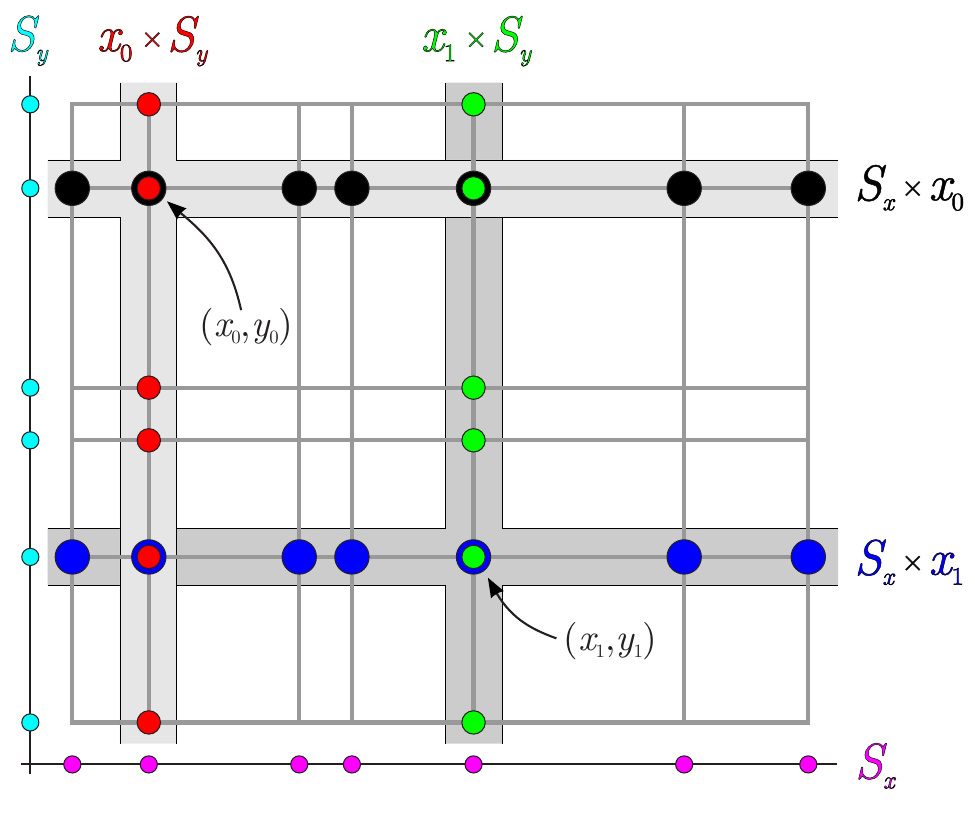}%
}{%
\begin{tikzpicture}[scale=0.9]
  \fill[gray!25] (-0.15,0) rectangle (0.15,4);      
  \fill[gray!25] (1.85,0) rectangle (2.15,4);       
  \fill[gray!25] (-1,2.85) rectangle (4,3.15);      
  \fill[gray!25] (-1,0.85) rectangle (4,1.15);      
  \draw[->] (-1.2,-0.3) -- (4.3,-0.3) node[right] {$S_x$};
  \draw[->] (-1.2,-0.3) -- (-1.2,4.3) node[above] {$S_y$};
  \fill (0,3) circle (2pt) node[above left] {$\langle x_0,y_0\rangle$};
  \fill (2,1) circle (2pt) node[below right] {$\langle x_1,y_1\rangle$};
  \draw (2,3) circle (2pt) node[above right] {$\langle x_1,y_0\rangle$};
  \node at (0,-0.6) {$x_0$}; \node at (2,-0.6) {$x_1$};
  \node at (-1.5,3) {$y_0$}; \node at (-1.5,1) {$y_1$};
\end{tikzpicture}%
}
\captionof{figure}{$f$ is constant on the sets within the shaded gray areas (the row and the column through each good point). Thus $f(x_0,y_0)$ must be equal to $f(x_1,y_1)$ as perpendicular lines cross.}
\label{fig1}
\end{center}

\noindent
\underline{Proof of Theorem \ref{thm:one}:}
(i) If $x_0 \in S \subseteq f^{-1}(f(x_0))$, then $x_0 \in S \subseteq (g \circ f)^{-1}(g(f(x_0)))$. Hence every set admissible for $f$ at $x_0$ is admissible for $g\circ f$ at $x_0$, and the minimum for $g \circ f$ is not larger. Moreover, by Proposition \ref{prop:collapse} applied to $g \circ f$ at $x_0$ and to $g$ at $f(x_0)$, both $\frac{d g(f(x))}{dx} |_{x_0}$ and $\frac{d g(y)}{dy}|_{f(x_0)}$ equal $K(g(f(x_0)))$ up to an additive constant.

(ii) By assumption, there are two sets $S_x \ni x_0$ and $S_y \ni y_0$ with complexity less than $c$ such that $f(\{x_0\} \times S_y) = f(S_x \times \{y_0\}) = \{ f(x_0,y_0) \}$. In particular, $\langle x_0, y_0\rangle$ is good for $f$, $S_x$, $S_y$. There are fewer than $2^c$ strings of length less than $c$, hence fewer than $2^{c}$ indices $i$ with $K(i) < c$, and fewer than $2^{2c}$ pairs of such indices. For each such pair $(i,j)$ for which $\varphi_i$, $\varphi_j$ are total and the corresponding product $S_i \times S_j$ contains a good point, let $V(i,j)$ denote the common value of $f$ at its good points (Lemma \ref{lem:good}). This defines a finite set of values $\mathcal{V}_c$, depending only on $c$ and $f$, and $v = f(x_0,y_0) \in \mathcal{V}_c$. Let $c_0 = \max \{ K(w) \,|\, w \in \mathcal{V}_c\}$. By Proposition \ref{prop:collapse} applied to $\tilde f$, $\partial_z f \,|_{\langle x_0,y_0\rangle} \le K(v) + c_1 \le c_0 + c_1$, where $c_1$ depends only on $f$. Put $c' = c_0 + c_1 + 1$.
$\qedsymbol{}$

\vspace{0.2cm}
The constant $c'$ obtained in this proof exists, but the proof gives no way to compute it from $c$, and there is a reason for this. To decide whether a point is good one must check that $f$ is constant on an infinite row and an infinite column, a $\Pi^0_1$ condition. This is where the halting problem enters. The precise statements are as follows. The first shows that with the halting problem as an oracle, the argument of Fig. \ref{fig1} is effective.

\begin{proposition}\label{prop:relativized}
Let $K^{0'}$ denote prefix complexity relative to the halting problem $0'$. For every computable $f(x,y)$, for all $x_0,y_0$,
\[
K^{0'}(f(x_0,y_0)) \lep \partial_x f \,|_{x_0,y_0} + \partial_y f \,|_{x_0,y_0} \quad .
\]
\end{proposition}
\begin{proof}
Let $i$, $j$ be indices of witness sets $S_x$, $S_y$ attaining $\partial_x f$ and $\partial_y f$. Given $i^*$ and $j^*$ (concatenated; this is self-delimiting), an oracle machine computes $i$ and $j$, enumerates the pairs $\langle x_1,y_1\rangle \in S_x \times S_y$ in some fixed order (membership is decidable), and for each pair asks the oracle whether it is good. ``Not good'' is witnessed by a finite search (an $x \in S_x$ or a $y \in S_y$ at which $f$ differs from $f(x_1,y_1)$), so the set of non-good pairs is recursively enumerable and goodness is decidable relative to $0'$. The search terminates, because $\langle x_0, y_0\rangle$ is good. The machine outputs $f$ at the first good pair found, which equals $f(x_0,y_0)$ by Lemma \ref{lem:good}. Hence $K^{0'}(f(x_0,y_0)) \lep K(i) + K(j)$.
\end{proof}

Without the oracle, the constant $c'$ in Theorem \ref{thm:one}(ii) cannot in general be bounded by a computable function of $c$. Let $B(k) \equiv \max \{ n \,|\, K(n) \le k \}$. It is standard that $B$ eventually exceeds every computable function: if $g$ is computable and increasing, then $K(g(k)) \lep K(k) \lep 2\log k$, so $K(g(k)) \le k$, and hence $g(k) \le B(k)$, for all large $k$.

\begin{proposition}\label{prop:busybeaver}
There exist a computable $f(x,y)$ and constants $d_0$, $d_1$ such that, for every $n$, there are $x_0 = y_0$ with
\[
\partial_x f \,|_{x_0,y_0} \le K(n) + d_0 \;, \quad \partial_y f \,|_{x_0,y_0} \le K(n) + d_0 \;, \quad \partial_z f \,|_{\langle x_0,y_0\rangle} \ge n - K(n) - d_1 \;.
\]
Consequently, any admissible $c'$ in Theorem \ref{thm:one}(ii) satisfies $c'(c) \ge B(c - d_0 - 1) - c - d_1$, which eventually exceeds every computable function of $c$.
\end{proposition}
\begin{proof}
Let $h_n(s)$ be the number of programs $p$ with $l(p) \le n$ such that $U(p)$ halts within $s$ steps. Then $h_n(s)$ is computable from $\langle n,s\rangle$ and non-decreasing in $s$. Since there are finitely many programs of length $\le n$, there is a least $t^*_n$ such that $h_n(s) = N_n$ for all $s \ge t^*_n$, where $N_n$ is the number of halting programs of length $\le n$. Define
\[
f(\langle m,i\rangle, \langle m',j\rangle) =
\begin{cases} h_m(\max(i,j)) & \text{if } m = m' \\ 0 & \text{otherwise,} \end{cases}
\]
which is total computable, and put $x_0 = y_0 = \langle n, t^*_n\rangle$. Then $f(x_0, y_0) = N_n$.

Let $S = \{ \langle n, i\rangle \,|\, i \in \N\}$. An index of $S$ is computable from $n$, so $K(S) \le K(n) + d_0$ for a constant $d_0$. For every $\langle n,i \rangle \in S$, $f(\langle n,i\rangle, y_0) = h_n(\max(i, t^*_n)) = N_n$, so $S$ is admissible in (\ref{eq:weak}) for $\partial_x f$, and by symmetry for $\partial_y f$. This gives the first two inequalities.

For the third, by Proposition \ref{prop:collapse} it suffices to show $K(N_n) \ge n - K(n) - O(1)$. Given $n$ and $N_n$, run all programs of length $\le n$ in parallel until $N_n$ of them have halted; these are then all the halting programs of length $\le n$. Output the least string $w$ that is not the output of any of them. Then $K(w) > n$, since otherwise a program of length $K(w) \le n$ would output $w$. On the other hand $K(w) \lep K(\langle n, N_n\rangle) \lep K(n) + K(N_n)$ by (F2). Hence $K(N_n) \ge n - K(n) - O(1)$.

For the consequence: take $n = B(c - d_0 - 1)$, so that $K(n) \le c - d_0 - 1$ and both partial derivatives are less than $c$. Then $c' > \partial_z f \ge n - K(n) - d_1 \ge B(c-d_0-1) - c - d_1$.
\end{proof}

In the example of Proposition \ref{prop:busybeaver} the two arguments coincide, $x_0 = y_0$, so they share all their information. This is unavoidable. The following theorem is the effective form of Theorem \ref{thm:one}(ii): the only obstruction is the mutual information between the arguments.

\begin{lemma}\label{lem:key}
For every computable $f(x,y)$ there is a constant $c$ such that for all $x_0, y_0$, with $v = f(x_0,y_0)$,
\[
K(v) \le K(v \,|\, x_0) + K(v \,|\, y_0) + I(x_0 : y_0) + c \quad .
\]
\end{lemma}
\begin{proof}
All (in)equalities below hold up to additive constants depending only on $f$ and $U$. Since $v$ is computable from $\langle x_0, y_0\rangle$, (F2) and (F3) give
\begin{align*}
K(\langle x_0, y_0 \rangle) &\eqp K(\langle v, \langle x_0, y_0 \rangle \rangle) \eqp K(v) + K(\langle x_0, y_0 \rangle \,|\, v^*) \\
&\lep K(v) + K(x_0 \,|\, v^*) + K(y_0 \,|\, v^*) \quad .
\end{align*}
By (F3) applied twice, $K(v) + K(x_0 \,|\, v^*) \eqp K(\langle v, x_0 \rangle) \eqp K(x_0) + K(v \,|\, x_0^*) \lep K(x_0) + K(v \,|\, x_0)$, i.e. $K(x_0 \,|\, v^*) \lep K(x_0) + K(v \,|\, x_0) - K(v)$, and similarly $K(y_0 \,|\, v^*) \lep K(y_0) + K(v \,|\, y_0) - K(v)$. Substituting,
\[
K(\langle x_0, y_0 \rangle) \lep K(x_0) + K(y_0) + K(v \,|\, x_0) + K(v \,|\, y_0) - K(v) \quad ,
\]
which is the claim after rearranging, since $I(x_0 : y_0) = K(x_0) + K(y_0) - K(\langle x_0,y_0\rangle)$.
\end{proof}

\begin{theorem}[Effective form of Theorem \ref{thm:one}(ii)]\label{thm:effective}
For every computable $f(x,y)$, for all $x_0, y_0$,
\begin{align*}
\partial_z f \,|_{\langle x_0,y_0\rangle} \; &\lep \; \partial_x f \,|_{x_0,y_0} + \partial_y f \,|_{x_0,y_0} + I(x_0 : y_0) \quad , \\
\partial_z f \,|_{\langle x_0,y_0\rangle} \; &\lep \; \partial^{\,|}_x f \,|_{x_0,y_0} + \partial^{\,|}_y f \,|_{x_0,y_0} + I(x_0 : y_0) \quad .
\end{align*}
In particular, if $I(x_0:y_0) \le d$, then $\partial_{x,y} f \,|_{x_0,y_0} < c$ implies $\partial_z f \,|_{\langle x_0,y_0\rangle} < 2c + d + c''$, with $c''$ depending only on $f$.
\end{theorem}
\begin{proof}
By Proposition \ref{prop:collapse} (for $\tilde f$), $\partial_z f \,|_{\langle x_0,y_0\rangle} \eqp K(v)$. By Proposition \ref{prop:partial}, $K(v\,|\,y_0)$ is bounded above, up to a constant, both by $\partial_x f$ and by $\partial^{\,|}_x f$, and similarly $K(v\,|\,x_0)$ by $\partial_y f$ and by $\partial^{\,|}_y f$. Now apply Lemma \ref{lem:key}.
\end{proof}

\begin{remark}\label{rem:condfails}
For the conditional variant (\ref{eq:weakcond}) the term $I(x_0:y_0)$ cannot be dropped even non-effectively. Take $f(x,y) = x$ and $x_0 = y_0$. Then $\partial^{\,|}_x f \eqp K(x_0 \,|\, y_0) \eqp 0$ and $\partial^{\,|}_y f \eqp K(x_0\,|\,x_0) \eqp 0$, while $\partial_z f \eqp K(x_0)$ is unbounded. For the unconditional definition (\ref{eq:weak}), Theorem \ref{thm:one}(ii) does hold, but by Proposition \ref{prop:busybeaver} only with a non-computable $c'$.
\end{remark}

\subsection{Weak dependence and the conditional independence axioms}\label{sec:ci}

In this section we show that the conditional independence axioms (C1)--(C5) hold for algorithmic conditional mutual information, with explicit losses, and that the results of the previous sections are instances of them. The idea of replacing stochastic conditional independence with the vanishing of algorithmic conditional mutual information is due to Janzing and Sch\"olkopf \cite{janzing2010}. That such relations satisfy the semi-graphoid axioms (C1)--(C4) is a general property of submodular information measures \cite{steudel2010}. We give self-contained proofs, with the losses made explicit.

\begin{definition}[Algorithmic conditional independence]\label{def:ci}
For $k \ge 0$ and strings $x,y,z$, we write $x \independent_k y \,|\, z$ if $I(x:y\,|\,z) \le k$, and $x \independent_k y$ if $I(x:y) \le k$. Pairs of strings are treated as strings via the pairing function; ``$(y,z)$'' stands for $\pair{y,z}$.
\end{definition}

Thus $x \independent_k y \,|\, z$ means that, given $z$, knowing $y$ shortens the description of $x$ by at most $k$ bits (up to logarithmic terms, by (F4)). The following chain rule is the basic tool.

\begin{lemma}[Chain rule]\label{lem:chain}
There is a constant $c$ such that for all strings $x,y,w,z$, with $n = l(x) + l(y) + l(w) + l(z)$ and $\delta_n = c \log (n+2)$,
\[
\bigl| \, I(x : \pair{y,w} \,|\, z) - I(x:y\,|\,z) - I(x : w \,|\, \pair{y,z}) \, \bigr| \le \delta_n \quad .
\]
\end{lemma}
\begin{proof}
By (F4), each of the following holds up to $O(\log(n+2))$ (the lengths of the pairs involved are bounded by a constant multiple of $n$ plus a constant):
\begin{align*}
K(\pair{x,y} \,|\, z) &= K(y\,|\,z) + K(x \,|\, \pair{y,z}) \quad , \\
K(\pair{y,w} \,|\, z) &= K(y\,|\,z) + K(w \,|\, \pair{y,z}) \quad , \\
K(\pair{x,\pair{y,w}} \,|\, z) &= K(y\,|\,z) + K(\pair{x,w} \,|\, \pair{y,z}) \quad ,
\end{align*}
where in the last line we also used that $\pair{x,\pair{y,w}}$ and $\pair{y,\pair{x,w}}$ are computable from each other. Expanding the definitions,
\begin{align*}
& I(x:y\,|\,z) + I(x:w\,|\,\pair{y,z}) \\
& \quad = K(x\,|\,z) + K(y\,|\,z) - K(\pair{x,y}\,|\,z) \\
& \qquad + K(x\,|\,\pair{y,z}) + K(w\,|\,\pair{y,z}) - K(\pair{x,w}\,|\,\pair{y,z}) \; .
\end{align*}
Substituting the first identity, $K(y\,|\,z)$ and $K(x\,|\,\pair{y,z})$ cancel, and the right-hand side becomes $K(x\,|\,z) + K(w\,|\,\pair{y,z}) - K(\pair{x,w}\,|\,\pair{y,z})$. By the second and third identities this equals $K(x\,|\,z) + K(\pair{y,w}\,|\,z) - K(\pair{x,\pair{y,w}}\,|\,z) = I(x : \pair{y,w} \,|\, z)$.
\end{proof}

\begin{theorem}[Algorithmic semi-graphoid]\label{thm:semigraphoid}
There is a constant $c$ (for (C2) and (C3), depending also on the computable function $h$) such that, for all strings $x,y,z,w$, with $n$ the sum of their lengths and $\delta_n = c\log(n+2)$:
\begin{enumerate}[label=(C\arabic*$'$)]
\item $x \independent_k y \,|\, z \implies y \independent_{k+c} x \,|\, z$;
\item $x \independent_k y \,|\, z$ and $u=h(x) \implies u \independent_{k+\delta_n} y \,|\, z$;
\item $x \independent_k y \,|\, z$ and $u=h(y) \implies x \independent_{k+\delta_n} y \,|\, (z,u)$;
\item $x \independent_k y \,|\, z$ and $x \independent_l w \,|\, (y,z) \implies x \independent_{k+l+\delta_n} (w,y) \,|\, z$.
\end{enumerate}
\end{theorem}
\noindent The proof, which uses only Lemma \ref{lem:chain} and (F2), is given in Appendix \ref{app:proofs}.

\begin{theorem}[Algorithmic intersection property]\label{thm:intersection}
There is a constant $c$ such that for all strings $x,y,z$, with $\delta_n$ as above,
\[
I(x : \pair{y,z}) \le I(x:y\,|\,z) + I(x:z\,|\,y) + I(y:z) + \delta_n \quad .
\]
Thus, (C5$'$): $x \independent_k y \,|\, z$, $x \independent_l z \,|\, y$ and $y \independent_m z$ imply $x \independent_{k+l+m+\delta_n} (y,z)$.
\end{theorem}
\begin{proof}
Apply Lemma \ref{lem:chain} with the empty condition (note $K(\cdot\,|\,\varepsilon) \eqp K(\cdot)$). First, $I(x:\pair{y,z}) \le I(x:y) + I(x:z\,|\,y) + \delta_n$. Second, expanding $I(y : \pair{x,z})$ in two ways,
\begin{align*}
I(y:z) + I(y:x\,|\,z) &\ge I(y:\pair{z,x}) - \delta_n \ge I(y:\pair{x,z}) - c - \delta_n \\
&\ge I(y:x) + I(y:z\,|\,x) - c - 2\delta_n \ge I(y:x) - 2c - 2\delta_n \quad ,
\end{align*}
so $I(x:y) \le I(y:z) + I(x:y\,|\,z) + O(1) + 2\delta_n$ (using symmetry). Combining the two inequalities and enlarging $c$ gives the claim.
\end{proof}

The third hypothesis $y \independent_m z$ in (C5$'$) is the quantitative form of the proviso that ``there is no non-trivial relationship between $Y$ and $Z$''. In the setting of weak dependence the logarithmic term $\delta_n$ can be dispensed with: this is Lemma \ref{lem:key}, which is the case $x = v = f(y,z)$ of Theorem \ref{thm:intersection}, proved with $O(1)$ precision.

\begin{corollary}[Weak dependence as conditional independence]\label{cor:dictionary}
Let $f$ be computable and $v = f(x_0,y_0)$. Up to additive constants depending only on $f$:
\begin{enumerate}[label=(\roman*)]
\item $f'(x)\,|_{x_0} \eqp I(x_0 : f(x_0))$, so ``$f$ depends weakly on $x$ at $x_0$'' means $f(x_0) \independent_k x_0$ for small $k$;
\item $\partial^{\,|}_x f\,|_{x_0,y_0} \eqp I(x_0 : v \,|\, y_0)$, so ``$f$ depends weakly on $x$ at $(x_0,y_0)$, given $y_0$'' means $v \independent_k x_0 \,|\, y_0$ for small $k$;
\item Theorem \ref{thm:one}(i) is an instance of (C2): $g(v) = g \circ f (x_0)$ is a computable function of $v$, and $I(x_0 : g(v)) \lep I(x_0 : v)$;
\item Theorem \ref{thm:one}(ii) and Theorem \ref{thm:effective} are instances of (C5): $v \independent x_0 \,|\, y_0$ and $v \independent y_0 \,|\, x_0$ imply $v \independent (x_0, y_0)$, provided $x_0 \independent y_0$.
\end{enumerate}
\end{corollary}
\begin{proof}
(i) is Corollary \ref{cor:mi}, (ii) is Proposition \ref{prop:partial}(ii), (iii) follows from Theorem \ref{thm:one}(i) and Corollary \ref{cor:mi}, and (iv) is Theorem \ref{thm:effective} together with $I(\pair{x_0,y_0} : v) \eqp K(v) \eqp \partial_z f\,|_{\pair{x_0,y_0}}$.
\end{proof}

\subsection{Kolmogorov complexity as the canonical notion of determination}\label{sec:determination}

In Section 3, statements of the form ``$y$ is determined by $x$ up to $k$ bits'' are used to recognize copies, to express that offspring inherit perturbations, and to express that machines can be decoded from code. Copies may be encoded in any computably recoverable form, so the notion of determination should not depend on the particular encoding. We now show that, under natural requirements, this notion is given by conditional Kolmogorov complexity, up to logarithmic terms.

A determination system is a family of relations $D_k \subseteq \{\text{strings}\}^2$, $k \in \N$, where $D_k(x,y)$ is read ``$y$ is determined by $x$ up to $k$ bits''. Write $K_D(y \,|\, x) = \min \{ k \,|\, D_k(x,y)\}$. For a partial computable function $g$, let $K(g) = \min\{K(i) \,|\, \varphi_i = g\}$. Consider the following requirements.
\begin{enumerate}[label=(E\arabic*)]
\item \emph{Effectivity:} the set $\{ \langle x, y, k\rangle \,|\, D_k(x,y)\}$ is recursively enumerable. (A determination can be certified.)
\item \emph{Counting:} there is a constant $a$ such that for all $x$ and $k$, at most $2^{k+a}$ strings $y$ satisfy $D_k(x,y)$. (Few things are determined by few bits.)
\item \emph{Re-encoding:} there is a constant $a'$ such that $D_{K(g) + a'}(x, g(x))$ holds for every partial computable $g$ and every $x$ at which $g$ is defined. (Applying a computable re-encoding costs no more than describing it.)
\end{enumerate}
The relations $D^K_k(x,y) :\iff K(y\,|\,x) \le k$ satisfy (E1)--(E3): the set in (E1) is enumerated by running all programs in parallel, (E2) holds with $a = 1$ since there are fewer than $2^{k+1}$ programs of length at most $k$, and (E3) holds by (F2), since $g(x)$ is computed from $x$ by a program for $g$.

\begin{theorem}[Characterization of determination]\label{thm:determination}
Let $D$ be a determination system.
\begin{enumerate}[label=(\roman*)]
\item If $D$ satisfies (E1) and (E2), then there is a constant $c_D$ such that for all $x$, $y$,
\[
K(y \,|\, x) \le K_D(y \,|\, x) + 2 \log_2 \bigl(K_D(y\,|\,x) + 2\bigr) + c_D \quad .
\]
\item If $D$ satisfies (E3), then there is a constant $c$ such that for all $x$, $y$,
\[
K_D(y \,|\, x) \le K(y\,|\,x) + 2 \log_2 \bigl(K(y\,|\,x) + 2\bigr) + c \quad .
\]
\end{enumerate}
Hence every determination system satisfying (E1)--(E3) agrees with conditional Kolmogorov complexity up to logarithmic terms.
\end{theorem}

\begin{proof}
(i) Let $k = K_D(y\,|\,x)$. By (E1), the set $S_{x,k} = \{ y' \,|\, D_k(x,y')\}$ can be enumerated uniformly in $x$ and $k$, and by (E2) it has at most $2^{k+a}$ elements. Hence $y$ is determined by $x$, by $k$, and by the position of $y$ in the enumeration of $S_{x,k}$, which is a number below $2^{k+a}$ and can be written with exactly $k + a$ bits once $k$ is known. A self-delimiting description of $k$ takes at most $2\log_2(k+2) + O(1)$ bits, so $K(y\,|\,x) \le k + a + 2 \log_2(k+2) + O(1)$.
(ii) Let $p$ be a shortest program with $U(p, x) = y$, so $l(p) = K(y\,|\,x)$, and let $g_p = U(p, \cdot)$, a partial computable function with $g_p(x) = y$. An index of $g_p$ is computable from $p$ (by the $s$-$m$-$n$ theorem), so $K(g_p) \le K(p) + O(1) \le l(p) + 2\log_2(l(p)+2) + O(1)$. By (E3), $K_D(y\,|\,x) \le K(g_p) + a'$.
\end{proof}

Part (i) says that Kolmogorov complexity is, up to logarithmic terms, the finest effective notion of determination that respects counting: any other such notion certifies only determinations that $K$ also certifies. Part (ii) says that a notion that charges computable re-encodings no more than their description length certifies all of them. Consequently, the upper-bound conditions of Section 3, which must accept copies in any computably recoverable form, can be stated with $K$ without loss of generality, up to logarithmic slack in the parameter $k$. The requirements (E1)--(E3) are the familiar ingredients of the invariance and optimality theorems for Kolmogorov complexity \cite{li2008}; the point here is that they are forced by the intended use. Note that (E3) refers to the description length $K(g)$ of transformations. The theorem therefore does not derive Kolmogorov complexity from nothing. What it shows is that a notion of determination between objects that is effective, respects counting, and is compatible with measuring transformations by their description length must agree with Kolmogorov complexity on objects as well.

Further material on these notions is collected in Appendix \ref{app:further}: an abstract formulation in terms of Dawid's separoids, the exact probabilistic counterpart of the ``perpendicular lines cross'' argument of Fig. \ref{fig1} together with its relation to common information, and a precise sense in which $f'(x)\,|_{x_0}$ and $\partial_x f$ behave like derivatives (which motivates the notation).

\section{Application to self-reproducing machines.}\label{sec:selfrep}
Consider a cellular automaton, akin to that in the construction of von Neumann and Burks, harboring a self-replicating machine. Since the details of implementation will soon become irrelevant, we will initially remain intentionally vague on those details. Any notion of self-reproduction within this context requires two distinctions. The first is between the object that reproduces and its environment, together with the dynamical law. The second is, within the object, between the part that is copied (the description, or code) and the part that does the work (the machine).

We express both distinctions using two ingredients only: algorithmic information, measured relative to the dynamical law, and interventions, i.e. changes of the initial configuration. We proceed as follows.
\begin{enumerate}[label=(\alph*)]
\item We express the content of a region of cellular space at time $t$ as a computable function of an initial ``patch'' and its environment, so that the causal quantities of Section \ref{sec:ci} apply (Sections \ref{sec:ca}--\ref{sec:patches}).
\item We identify copies of an object by the information they share with it, rather than by a predefined pattern, and show that proliferation of copies amounts to unbounded growth of shared information (Section \ref{sec:copies}).
\item We identify the code and the machine of an object by their response to perturbations: perturbations of the code are inherited by the offspring, perturbations of the machine are not (Section \ref{sec:heritability}).
\item We define self-reproduction (Section \ref{sec:selfrepdef}), and measure its amount in two ways: algorithmically, by the complexity relative to the rule of the simplest machinery that can stand in for the object's machine, and by certification, as the length of the shortest proof, in a fixed formal theory, that some machinery can stand in for it (Section \ref{sec:certified}).
\item We show, in an abstract model of reproduction, that neither ingredient can be dispensed with: criteria computed from the actual run alone, including all algorithmic information quantities, cannot distinguish a machine that is built from its description from one that is copied; and criteria that do not condition on the rule attribute to the object information that the rule supplies (Section \ref{sec:necessity}).
\item We reduce the claim that the definition includes the construction of von Neumann and Burks to explicit hypotheses (Section \ref{sec:vonneumann}).
\end{enumerate}
The qualitative definition uses only counts, interventions and upper bounds on information. This matters: a requirement that some quantity of information be \emph{large} can be undermined by exhaustive search, since a replicator found by dovetailing through all patterns has a short description given the dynamical law. The algorithmic measure of the amount of self-reproduction is a lower-bound quantity of this kind, and is affected by search (Remark \ref{rem:search}). The certified measure is not: a search can verify that a machine has produced $N$ copies by time $T$, but it cannot produce a proof that copies are produced for every $N$, so it cannot shorten such a proof.

\subsection{Cellular automata}\label{sec:ca}

Fix a dimension $d \ge 1$, a finite alphabet of states $\Sigma$ containing a distinguished quiescent state $0$, a radius $\rho \ge 1$, and a local rule $\phi : \Sigma^{N_\rho} \to \Sigma$, where $N_\rho = \{-\rho, \ldots, \rho\}^d$, satisfying $\phi(0,\ldots,0) = 0$. A configuration is a map $c : \mathbb{Z}^d \to \Sigma$; its support is $\operatorname{supp}(c) = \{ i \,|\, c(i) \ne 0\}$, and $c$ is finite if its support is finite. The global map is
\[
F(c)(i) = \phi\bigl( (c(i+e))_{e \in N_\rho} \bigr) \quad , \qquad i \in \mathbb{Z}^d \quad .
\]
Since $\phi(0,\ldots,0)=0$, $F$ maps finite configurations to finite configurations. We write $c_t = F^t(c_0)$ and call the sequence $(c_t)_{t \ge 0}$ the run of $c_0$.

We use the $\ell_\infty$ distance $|i - j|_\infty$ on $\mathbb{Z}^d$. For $R \subseteq \mathbb{Z}^d$ and $r \ge 0$ let $R^{+r} = \{ i \,|\, |i-j|_\infty \le r \text{ for some } j \in R\}$. Two cells are adjacent if they are at distance $1$, and a set of cells is connected if it is connected with respect to adjacency. A box is a set $\prod_{l=1}^d \{a_l, \ldots, b_l\}$, and $\operatorname{box}(R)$ denotes the smallest box containing a finite set $R$. A pattern is a pair $(D, \pi)$ with $D \subseteq \mathbb{Z}^d$ finite and $\pi : D \to \Sigma$; $c|_D$ denotes the restriction of $c$ to $D$, and $\operatorname{supp}(\pi) = \{ i \in D \,|\, \pi(i) \neq 0\}$. Boxes, the local rule $\phi$ (given as a table), finite configurations, and the restrictions of configurations to boxes (given as arrays together with the position of the box) are finite objects, and we fix computable encodings of them by strings; all complexities below refer to these encodings.

\begin{lemma}[Light cone]\label{lem:cone}
If $c$ and $c'$ agree on $B^{+\rho t}$, then $F^t(c)$ and $F^t(c')$ agree on $B$.
\end{lemma}
\begin{proof}
For $t=0$ there is nothing to prove. If $c$ and $c'$ agree on $B^{+\rho (t+1)} = (B^{+\rho})^{+\rho t}$, then by induction $F^t(c)$ and $F^t(c')$ agree on $B^{+\rho}$. Since $F(\cdot)(i)$ depends only on the values on $i + N_\rho \subseteq B^{+\rho}$ for $i \in B$, $F^{t+1}(c)$ and $F^{t+1}(c')$ agree on $B$.
\end{proof}

\emph{Interventions.} In a cellular automaton, an intervention in the sense of Pearl's do-operator \cite{pearl2009} is simply a change of the initial configuration. For a finite configuration $c$ and a set of cells $R$, we write $c[R := 0]$ for the configuration that coincides with $c$ outside $R$ and is quiescent on $R$; more generally, for a set of states $\Lambda \subseteq \Sigma$, $c[R \cap \Lambda := 0]$ denotes the configuration obtained from $c$ by replacing with $0$ every cell of $R$ whose state lies in $\Lambda$.

\subsection{Causation between patches of cells}\label{sec:patches}

A frame is a tuple $\theta = \langle \phi, t, A, B\rangle$, where $A$ and $B$ are boxes and $t \ge 0$; $A$ is the ``patch'' at time $0$ and $B$ the region observed at time $t$. Let $E = B^{+\rho t} \setminus A$ (the part of the light cone of $B$ outside the patch). For a finite configuration $c_0$ put
\[
x = c_0|_A \quad , \qquad y = c_0|_E \quad , \qquad v = c_t|_B \quad .
\]
By Lemma \ref{lem:cone}, $v$ is determined by $\theta$, $x$ and $y$: it equals $F^t(\hat c)|_B$, where $\hat c$ coincides with $x$ on $A$, with $y$ on $E$, and is $0$ elsewhere. Hence the function
\[
f(x, \langle \theta, y \rangle) = F^t(\hat c)|_B
\]
(defined to be $0$ on inputs that are not of this form) is a total computable function of two arguments, and $v = f(x, \langle \theta, y\rangle)$. Note that the rule $\phi$ is part of the frame, so $f$ does not depend on the automaton. Consequently, all constants below depend only on the universal machine $U$ and on the chosen encodings, and not on the automaton.

\begin{definition}[Causal strength and screening]\label{def:strength}
For a frame $\theta$ and a finite configuration $c_0$, with $x$, $y$, $v$ as above, the causal strength of the patch $A$ at time $0$ on the region $B$ at time $t$, and the screening defect, are
\[
\kappa_\theta(c_0) \equiv K(v \,|\, \langle \theta, y \rangle) \quad , \qquad \sigma_\theta(c_0) \equiv K(v \,|\, \langle \theta, x \rangle) \quad .
\]
\end{definition}

In words: $\kappa$ is the number of bits of the content of $B$ at time $t$ that cannot be obtained from the environment of the patch (and the frame), and $\sigma$ is the number of bits of that content that cannot be obtained from the patch. ``The patch $A$ causes the content of $B$'' is expressed by $\sigma$ being small: the content of $B$ is, up to few bits, independent of everything else given the patch (compare the informal description following (C1)--(C5)).

\begin{proposition}\label{prop:strength-ci}
Up to additive constants depending only on $U$,
\[
\kappa_\theta(c_0) \eqp \partial^{\,|}_x f \,\big|_{x, \langle \theta, y \rangle} \eqp I\bigl(x : v \,|\, \langle \theta, y\rangle\bigr) \quad , \qquad
\sigma_\theta(c_0) \eqp I\bigl(y : v \,|\, \langle \theta, x\rangle\bigr) \quad .
\]
\end{proposition}
\begin{proof}
The first two relations are Proposition \ref{prop:partial}(ii) applied to the computable function $f$ at the point $(x, \langle \theta, y\rangle)$. For the third, note that $v$ is computable from $y$ given $\langle \theta, x\rangle$, so $K(\langle y, v\rangle \,|\, \langle \theta, x \rangle) \eqp K(y \,|\, \langle \theta, x \rangle)$ by (F2), whence $I(y : v \,|\, \langle \theta, x\rangle) \eqp K(v \,|\, \langle \theta, x\rangle)$.
\end{proof}

Thus $\kappa$ and $\sigma$ are the two conditional partial derivatives of the map ``initial configuration $\mapsto$ content of $B$ at time $t$'' with respect to the patch and to its environment. The following proposition, an instance of the intersection property (Theorem \ref{thm:intersection}), shows that together they account for the whole content of $B$, up to the information that the patch shares with its environment.

\begin{proposition}\label{prop:decomposition}
There is a constant $c$ such that, with $n$ the total length of $\theta$, $x$, $y$ and $v$,
\[
K(v \,|\, \theta) \le \kappa_\theta(c_0) + \sigma_\theta(c_0) + I(x : y \,|\, \theta) + c \log(n+2) \quad .
\]
\end{proposition}
\begin{proof}
The proof of Theorem \ref{thm:intersection} uses only Lemma \ref{lem:chain}, the bound $I(\cdot:\cdot\,|\,\cdot) \ge -c$, and symmetry, all of which hold with an arbitrary condition. Hence it applies verbatim with all quantities conditioned on $\theta$:
\[
I(v : \langle x, y\rangle \,|\, \theta) \le I(v : x \,|\, \langle y,\theta \rangle) + I(v : y \,|\, \langle x,\theta \rangle) + I(x : y \,|\, \theta) + c \log(n+2) \quad .
\]
Since $v$ is computable from $\langle x, y\rangle$ given $\theta$, $I(v : \langle x,y\rangle \,|\, \theta) \eqp K(v\,|\,\theta)$, as in the proof of Proposition \ref{prop:strength-ci}. The two remaining terms are $\kappa$ and $\sigma$ by Proposition \ref{prop:strength-ci} (the order of the components of a pair in the condition changes complexities by $O(1)$).
\end{proof}

\begin{remark}\label{rem:emptyenv}
If the environment is empty, i.e. $c_0$ vanishes on $E$, then $y$ is computable from $\theta$, so $\kappa_\theta(c_0) \eqp K(v \,|\, \theta)$: all information in $B$ at time $t$ comes either from the patch or from the frame (which contains the rule). Environments containing ``building blocks'' (cf. the definition of self-reproduction used in soft-matter physics, recalled in the Introduction) are handled by the same definitions; the term $I(x:y\,|\,\theta)$ in Proposition \ref{prop:decomposition} then measures how much the patch resembles its surroundings.
\end{remark}

\subsection{Copies and proliferation}\label{sec:copies}

To speak of individual objects at all, we need a minimal notion of separation. A box $B$ is delimited in a configuration $c$ if $c|_B \neq 0$ and $c$ vanishes on $B^{+1} \setminus B$. This formalizes the informal requirement that ``machines'' are separated regions of cellular space. It is the only structural, non-informational notion used below; in particular, we do not use any predefined indicator of the kind of object to be detected.

\begin{lemma}\label{lem:delimited}
Let $c$ be a finite configuration. For every delimited box $B$, the set $\operatorname{supp}(c) \cap B$ is a non-empty union of connected components of $\operatorname{supp}(c)$. Consequently, the number of pairwise disjoint delimited boxes in $c$ is at most the number of connected components of $\operatorname{supp}(c)$.
\end{lemma}
\begin{proof}
$\operatorname{supp}(c) \cap B \neq \emptyset$ since $c|_B \neq 0$. If $i \in \operatorname{supp}(c) \cap B$ and $j \in \operatorname{supp}(c)$ is adjacent to $i$, then $j \in B^{+1}$; since $c(j) \neq 0$ and $c$ vanishes on $B^{+1} \setminus B$, $j \in B$. So $\operatorname{supp}(c) \cap B$ is closed under adjacency within $\operatorname{supp}(c)$, and pairwise disjoint delimited boxes contain pairwise disjoint non-empty unions of components.
\end{proof}

Throughout the rest of this section, a parent is a box $A$ that is delimited in the initial configuration $c_0$, and $x = c_0|_A$. For a box $B$ and a time $t$, we use the frame $\theta = \langle \phi, t, A, B\rangle$ and the notation $y$, $v$, $\kappa_\theta$, $\sigma_\theta$ of Section \ref{sec:patches}.

\begin{definition}[Descendants and proliferation]\label{def:descent}
Let $k \ge 0$. A box $B = A + s$ that is delimited in $c_t$ is a $k$-descendant of the parent $A$ at time $t$ if
\begin{enumerate}[label=(D\arabic*)]
\item $\sigma_\theta(c_0) \le k$ \hfill (the offspring is determined by the parent);
\item $\kappa_\theta(c_0) \ge K(v \,|\, \theta) - k$ \hfill (the environment alone does not account for the offspring);
\item $K(x \,|\, \langle \theta, v \rangle) \le k$ \hfill (the parent can be recovered from the offspring).
\end{enumerate}
The parent $k$-proliferates if for every $N$ there are a time $t$ and $N$ pairwise disjoint $k$-descendants at time $t$.
\end{definition}

(D1) says that, given the parent, the offspring is independent of everything else up to $k$ bits; (D2) bounds $K(v\,|\,\theta) - \kappa_\theta(c_0)$, which is, up to logarithmic terms, the information about $v$ that the environment provides beyond the frame; and (D3) says that no information of the parent is lost. All three are upper bounds on information. We restrict attention to offspring that are translates of the parent; rotated or reflected copies could be admitted in the same way.

Copies are thus recognized by the information they share with the parent. The following proposition shows that proliferation is the same as unbounded growth of the information shared among disjoint regions, as measured by the total correlation $\sum_i K(v_i) - K(\langle v_1, \ldots, v_N\rangle)$.

\begin{proposition}[Proliferation as shared information]\label{prop:tc}
There is a constant $c$ such that the following holds. Let $B_1, \ldots, B_N$ be pairwise disjoint $k$-descendants of the parent $A$ at time $t$, let $v_l = c_t|_{B_l}$, $\Theta = \langle \phi, t, A, B_1, \ldots, B_N\rangle$, and let $n$ be the total length of $\Theta$, $x$ and the $v_l$. Then
\[
\sum_{l=1}^N K(v_l \,|\, \Theta) - K(\langle v_1, \ldots, v_N\rangle \,|\, \Theta) \;\ge\; (N-1)\, K(x \,|\, \Theta) - N \bigl(2k + c \log(n+2)\bigr) \quad .
\]
\end{proposition}
\begin{proof}
Let $\theta_l = \langle \phi, t, A, B_l\rangle$, which is computable from $\Theta$ and $l$, so that conditioning on $\Theta$ instead of $\theta_l$ costs at most $K(l) + O(1) = O(\log (n+2))$ bits. By (D1) and (D3), $K(v_l \,|\, \langle \Theta, x\rangle) \le k + O(\log(n+2))$ and $K(x \,|\, \langle \Theta, v_l\rangle) \le k + O(\log (n+2))$. By (F4) applied twice, $K(v_l \,|\, \Theta) + K(x \,|\, \langle \Theta, v_l\rangle)$ and $K(x \,|\, \Theta) + K(v_l \,|\, \langle \Theta, x\rangle)$ both equal $K(\langle x, v_l\rangle \,|\, \Theta)$ up to $O(\log(n+2))$, so
\[
K(v_l \,|\, \Theta) \ge K(x \,|\, \Theta) - k - O(\log (n+2)) \quad .
\]
On the other hand, by (F2), concatenating a shortest program for $x$ given $\Theta$ with shortest programs for each $v_l$ given $\langle \Theta, x\rangle$ (and $N$ is computable from $\Theta$),
\begin{align*}
K(\langle v_1, \ldots, v_N \rangle \,|\, \Theta) &\le K(x \,|\, \Theta) + \sum_{l} K(v_l \,|\, \langle \Theta, x\rangle) + O(N) \\
&\le K(x\,|\,\Theta) + N\bigl(k + O(\log(n+2))\bigr) \quad .
\end{align*}
Subtracting gives the claim.
\end{proof}

Thus, if the parent carries more than $2k + c\log(n+2)$ bits of information given the frame, proliferation implies that the information shared among disjoint regions grows without bound. Copies of trivial patterns carry little information, and are correspondingly invisible to this measure.

\begin{proposition}[The infection automaton does not proliferate]\label{prop:infection}
Let $\Sigma = \{0,1\}$, $\rho = 1$, and $\phi$ the maximum of the states in $N_1$ (so that $\operatorname{supp}(c_{t+1}) = \operatorname{supp}(c_t)^{+1}$). Then for every finite $c_0$ and every $t$, the number of pairwise disjoint delimited boxes in $c_t$ is at most the number of connected components of $\operatorname{supp}(c_0)$. In particular, no parent in any finite configuration $k$-proliferates, for any $k$.
\end{proposition}
\begin{proof}
Let $Y$ be a finite set of cells. Every cell of $Y^{+1}$ is adjacent or equal to a cell of $Y$, so every connected component of $Y^{+1}$ contains a cell of $Y$, together with the component of $Y$ containing it. Hence $Y^{+1}$ has at most as many components as $Y$. By induction, $\operatorname{supp}(c_t)$ has at most as many components as $\operatorname{supp}(c_0)$, and Lemma \ref{lem:delimited} gives the first claim; the second follows.
\end{proof}

The next example shows that proliferation alone is not a reasonable notion of self-reproduction: there are rules under which every pattern proliferates, so that the copying is a property of the dynamics rather than of the object. The example is a rule usually attributed to Fredkin, with a vector-valued alphabet.

Let $d = 2$, $\rho = 1$, $\Sigma = \mathbb{F}_2^r$ for some $r \ge 1$ (with $0$ the zero vector), and
\[
F(c)(i) = \sum_{e \in N_0} c(i+e) \quad , \qquad N_0 = \{\pm e_1, \pm e_2\} \quad ,
\]
the sum taken in $\mathbb{F}_2^r$.

\begin{lemma}[The parity rule copies every configuration]\label{lem:fredkin}
Let $c$ be a non-zero finite configuration, $A$ a box containing $\operatorname{supp}(c)$, and $\Delta$ the $\ell_\infty$-diameter of $A$. Let $0 \le k_1 < \ldots < k_J$ with $k_{j+1} \ge k_j + 3$ and $2^{k_1} \ge \Delta + 2$, and let $t = \sum_j 2^{k_j}$. Then there is a set $H \subseteq \mathbb{Z}^2$ of $4^J$ shifts such that the boxes $A - s$, $s \in H$, are pairwise disjoint and at distance at least $2$ from each other, and $F^t(c) = \sum_{s \in H} T_s c$, where $(T_s c)(i) = c(i+s)$. In particular, $F^t(c)|_{A - s}$ is the translate of $c|_A$ for every $s \in H$, and $F^t(c)$ vanishes outside $\bigcup_{s \in H} (A - s)$.
\end{lemma}

\begin{proof}
On finite configurations, viewed as vectors over the field $\mathbb{F}_2$, $F = \sum_{e \in N_0} T_e$ is linear and the $T_s$ commute. Hence $F^2 = \sum_{e,e'} T_{e+e'} = \sum_e T_{2e}$, because the terms with $e \ne e'$ occur in pairs $(e,e')$, $(e',e)$ and cancel modulo $2$. By induction, $F^{2^k} = \sum_{e \in N_0} T_{2^k e}$. Therefore
\[
F^t = \prod_{j=1}^J F^{2^{k_j}} = \sum_{(e_1, \ldots, e_J) \in N_0^J} T_{s(e_1,\ldots,e_J)} \quad , \qquad s(e_1,\ldots,e_J) = \sum_{j} 2^{k_j} e_j \quad .
\]
\emph{The $4^J$ shifts are pairwise distinct.} Let $(e_j) \neq (e'_j)$, let $j_0$ be the least index with $e_{j_0} \ne e'_{j_0}$, and let $w = e_{j_0} - e'_{j_0}$. Some coordinate $w_l$ of $w$ lies in $\{\pm1, \pm2\}$. The $l$-th coordinate of $s(e) - s(e')$ is $2^{k_{j_0}} w_l$ plus a multiple of $2^{k_{j_0}+3}$, and $0 < |2^{k_{j_0}} w_l| < 2^{k_{j_0}+3}$, so it is not zero. Hence no two terms cancel. Let $H$ be the set of these shifts.
\emph{Separation.} All coordinates of all shifts are multiples of $2^{k_1}$, so distinct shifts differ by at least $2^{k_1} \ge \Delta + 2$ in $\ell_\infty$ norm. Hence any cell of $A - s$ and any cell of $A - s'$, for $s \neq s'$, are at distance at least $2$. Since $T_s c$ is supported in $A - s$, the supports of the summands are disjoint, so their sum is their union, which gives the remaining claims.
\end{proof}

\begin{proposition}[A rule that copies every pattern]\label{prop:fredkin}
There is a constant $c$, depending only on $U$ and the encodings, such that for every non-zero finite configuration $c_0$ of the parity rule, the parent $A = \operatorname{box}(\operatorname{supp}(c_0))$ $c$-proliferates.
\end{proposition}
\begin{proof}
$A$ is delimited in $c_0$, since $c_0$ vanishes outside $A$. Let $t$ and $H$ be as in Lemma \ref{lem:fredkin}. For $s \in H$, the box $A - s$ is delimited in $c_t$: $c_t|_{A-s}$ is the translate of $c_0|_A \neq 0$, and the cells at distance $1$ from $A - s$ lie in no box $A - s'$, hence are quiescent. These $4^J$ boxes are pairwise disjoint. With $\theta = \langle \phi, t, A, A-s\rangle$, the content $v = c_t|_{A-s}$ is the same array as $x = c_0|_A$, so $v$ is computable from $\langle \theta, x\rangle$ and $x$ from $\langle \theta, v\rangle$, by a fixed program; this gives (D1) and (D3). The environment $y = c_0|_E$ is identically zero and computable from $\theta$, so $\kappa_\theta(c_0) \ge K(v \,|\, \theta) - c$, which is (D2). As $J$ is arbitrary, $A$ $c$-proliferates.
\end{proof}

In Proposition \ref{prop:fredkin} the parent may be arbitrarily complex (e.g. $c_0|_A$ may be an incompressible array), so no lower bound on the complexity of the parent helps. One might try to exclude this example by requiring that the ability to reproduce be a rare property of the object, e.g. by requiring a large ``functional information'' in the sense of Szostak and Hazen et al. \cite{szostak2003, hazen2007}, i.e. that a random pattern of the same size rarely reproduces. Rarity alone, however, does not distinguish an object that performs its own reproduction from one that merely carries a ``key'' recognized by the dynamics (see Proposition \ref{prop:keys} below). We take a different route, based on how the object responds to perturbations.

\subsection{Code and machine as heritable and non-heritable parts}\label{sec:heritability}

In the construction of von Neumann and Burks, as summarized in the Introduction, the copier $B$ acts on descriptions $d_X$, while machines are never copied: they are built by the constructor $A$ from descriptions. Consequently, a change in the description of the parent reappears in its offspring, while damage to the parent's machine does not: the offspring's machine is rebuilt from the (intact) description. This is the distinction, familiar from biology, between heritable changes of the germ line and non-heritable changes of the body (the Weismann barrier). We use it to define code and machine operationally, by interventions.

For a parent $A$ in $c_0$, a set of cells $R \subseteq A$, and an array $\delta \in \Sigma^R$, let $c_0^{R \leftarrow \delta}$ be the configuration that coincides with $\delta$ on $R$ and with $c_0$ elsewhere; it is a perturbation of the parent on $R$. Write $\varrho = c_0|_{A \setminus R}$ for the unperturbed rest of the parent. For a $k$-descendant $B$ at time $t$ (in the unperturbed run), let $v^\delta = c^{R\leftarrow\delta}_t|_B$ be the content of the same box at the same time in the perturbed run, and $\theta = \langle \phi, t, A, B\rangle$.

\begin{definition}[Heritable and somatic regions]\label{def:heritable}
Let $B$ be a $k$-descendant of $A$ at time $t$, and $R \subseteq A$.
\begin{enumerate}[label=(\roman*)]
\item $R$ is $k$-heritable at $(t,B)$ if, for at least half of all $\delta \in \Sigma^R$, $K(\delta \,|\, \langle \theta, \varrho, v^\delta\rangle) \le k$: the offspring region determines the perturbation.
\item $R$ is $k$-somatic at $(t,B)$ if, for at least half of all $\delta \in \Sigma^R$, $I(\delta : v^\delta \,|\, \langle \theta, \varrho\rangle) \le k$: the offspring region carries almost no information about the perturbation.
\end{enumerate}
\end{definition}

Both notions combine an intervention with an information measurement. A perturbation that destroys the machine, so that no offspring is formed, is not inherited: the offspring region then carries no information about it. Since both notions refer to most perturbations, they can be established by counting arguments, without proving that any particular string is complex. For more than half of all $\delta \in \Sigma^R$ we have $K(\delta \,|\, \langle \theta, \varrho\rangle) \ge |R| \log_2 |\Sigma| - 1$, by counting programs. For such a $\delta$, inheritance ($K(\delta \,|\, \langle \theta, \varrho, v^\delta\rangle) \le k$) gives $I(\delta : v^\delta \,|\, \langle \theta, \varrho\rangle) \ge |R| \log_2 |\Sigma| - k - O(\log(n+2))$ by (F4), which is far from the somatic bound $k$ when $R$ carries more than about $2k$ bits. In this sense the two notions are opposite.

\begin{definition}[Genotype and phenotype]\label{def:decomposition}
A decomposition of the parent $A$ is a partition $A = T \sqcup M$ with $c_0|_T \neq 0$ and $c_0|_M \neq 0$; we call $\gamma = c_0|_T$ the code and $m = c_0|_M$ the machine (arrays carrying their positions). The decomposition is a $k$-heritability decomposition if there is a $k$-descendant $B$ at some time $t$ such that $T$ is $k$-heritable and $M$ is $k$-somatic at $(t,B)$.
\end{definition}

Whether an object can be read as code together with machinery is thus a property to be verified by interventions, not an assumption. A decomposition is existentially quantified, but each of its parts has an operational meaning.

\begin{remark}[Linear typing]\label{rem:linear}
The heritable and somatic parts correspond to two kinds of types in linear logic \cite{girard1987}. Reusable types $!X$ admit duplication, $!X \multimap\, !X \otimes !X$ (contraction), while linear types do not. In this reading the copier $B$ is contraction on the type of descriptions, the constructor $A$ is a map from descriptions to machines, and the composite $D$ reproduces by duplicating its description and building a new machine from one of the copies. A machine that is never duplicated, only rebuilt from a description, is one whose perturbations are not inherited. Note also that typed theories have no internal operation producing the code of a given term; correspondingly, in the construction of von Neumann and Burks the description must be carried as data, rather than obtained by self-inspection. Appendix \ref{app:sorted} describes an alternative formulation in which code and machine are sorts of states, fixed in advance.
\end{remark}

\subsection{Self-reproduction}\label{sec:selfrepdef}

\begin{definition}[Self-reproduction]\label{def:selfrep}
Let $c, k \ge 0$. A parent $A$ in a finite configuration $c_0$, with a decomposition $A = T \sqcup M$, is a $(c,k)$-self-reproducer if:
\begin{enumerate}[label=(S\arabic*)]
\item \emph{(proliferation)} $A$ $k$-proliferates;
\item \emph{(heritability)} $T \sqcup M$ is a $k$-heritability decomposition;
\item \emph{(genotype determines phenotype)} $K\bigl(m \,\big|\, \langle \phi, A, \gamma\rangle\bigr) \le c$;
\item \emph{(agency)} after removing the machine, the code does not proliferate: with $c_0' = c_0[M := 0]$ and $x' = c_0'|_A$, there is a bound $N_0$ such that at no time $t$ are there $N_0$ pairwise disjoint boxes $B = A + s$, delimited in $c'_t$, with $K\bigl(x' \,\big|\, \langle \theta, c'_t|_B \rangle\bigr) \le k$.
\end{enumerate}
\end{definition}

Conditions (S1), (S2) and (S4) combine counts and interventions with upper bounds on information. Condition (S3) is an upper bound on information. (S3) says that the machine can be decoded from the code, up to $c$ bits. The constant $c$ is of the order of the complexity of the decoding procedure, the ``genetic code''.

\begin{proposition}[Exclusions]\label{prop:exclusions}\
\begin{enumerate}[label=(\roman*)]
\item In the infection automaton, no parent in any finite configuration is a $(c,k)$-self-reproducer, for any decomposition and any $c$, $k$.
\item In the parity rule (with any $r \ge 1$), there is a constant $k_0$ such that no parent in any finite configuration is a $(c,k)$-self-reproducer, for any decomposition, any $c$ and any $k \ge k_0$.
\end{enumerate}
\end{proposition}

\begin{proof}
(i) (S1) fails by Proposition \ref{prop:infection}.
(ii) We show that (S4) fails. Let $c_0' = c_0[M := 0]$ and $x' = c_0'|_A$; then $x' \neq 0$, since $c_0|_T \neq 0$. Let $A'' = \operatorname{box}(A \cup \operatorname{supp}(c_0'))$ and apply Lemma \ref{lem:fredkin} to $c_0'$ and the box $A''$. For every $s$ in the resulting set of $4^J$ shifts, the boxes $A'' - s$ are pairwise disjoint and at distance at least $2$ from each other, $c'_t|_{A'' - s}$ is the translate of $c'_0|_{A''}$, and $c'_t$ vanishes outside $\bigcup_s (A''-s)$. Hence $c'_t|_{A-s}$ is the translate of $x' \neq 0$. Moreover, $A - s$ is delimited in $c'_t$: a cell at distance $1$ from $A - s$ either lies in $A'' - s$, where $c'_t$ is the translate of $c'_0$, which vanishes on $A^{+1} \setminus A$ (because $A$ is delimited in $c_0$ and the intervention only changes cells of $M \subseteq A$), or lies outside $A'' - s$ at distance $1$ from it, hence in no box $A'' - s'$, where $c'_t$ vanishes. Finally, $x'$ is computable from $\langle \theta, c'_t|_{A-s}\rangle$ by a fixed program, so $K(x' \,|\, \langle \theta, c'_t|_{A-s}\rangle) \le k_0$ for a constant $k_0$. Since $J$ is arbitrary, (S4) fails for every $k \ge k_0$.
\end{proof}

In the parity rule, the code is copied whether or not the machine is present: the act of copying is performed by the dynamics and not by the object.

The definition is qualitative. To measure how much of the reproduction is performed by the object, we intervene on the content of the machine.

\begin{definition}[Algorithmic amount of self-reproduction]\label{def:amount}
For a self-reproducer as above, a machine variant is an array $m' \in \Sigma^M$; it is $k$-productive if, in the run of $c_0^{M \leftarrow m'}$, for every $N$ there are a time $t$ and $N$ pairwise disjoint boxes $B = A + s$, delimited in $c_t$, with $K(x \,|\, \langle \theta, c_t|_B\rangle) \le k$ and $K(c_t|_B \,|\, \langle \theta, x\rangle) \le k$, where $x = c_0|_A$ is the original parent. That is, the original object, with its original machine, proliferates, although the parent's machine has been replaced by $m'$. Let $W_k$ be the set of $k$-productive variants. The algorithmic amount of self-reproduction is
\[
\nu_k \equiv \min \bigl\{ K\bigl(m' \,\big|\, \langle \phi, A, M\rangle \bigr) \;\big|\; m' \in W_k \bigr\} \quad ,
\]
the complexity, given the rule, of the simplest machinery that can stand in for the parent's machine.
\end{definition}

The original machine is productive under the exact replication of Proposition \ref{prop:exact} below, and then $\nu_k \le K(m \,|\, \langle \phi, A, M\rangle)$. The amount is reported, not required to exceed a threshold.

\begin{remark}[Why the amount is measured on the simplest working machine]\label{rem:catalyst}
The following examples show why $\nu$ is defined as it is. They are described informally, since each requires a concrete rule.
\begin{enumerate}[label=(\alph*)]
\item \emph{Catalyst.} Suppose code cells are copied only next to a special machine state, and a new catalyst cell appears next to every completed copy. An object consisting of one catalyst cell and an arbitrary, possibly incompressible, string of code could satisfy (S1)--(S4). Measuring the amount by the information in the code would be misleading; since a single machine cell does the job, $\nu = O(1)$.
\item \emph{Transcription by the rule.} Suppose code is copied whenever some machine cell is adjacent, and the rule writes the offspring's machine from the code, cell by cell. The parent's machine may be large and complex, but a variant with a single machine cell adjacent to the code is productive, so again $\nu = O(1)$.
\item \emph{Inert machinery and keys.} If part of the machine plays no role in reproduction, variants that omit it are productive, so it does not contribute to $\nu$. If the rule copies code only next to a machine pattern equal to a fixed ``key'', the key is specified by the rule, so its complexity given $\phi$ is small (Proposition \ref{prop:keys}).
\item \emph{Checksums.} Suppose the rule copies code only next to a machine equal to a fixed computable function $g$ of the code, and writes $g(\text{code})$ as the offspring's machine. Then the simplest productive variant is $g(\gamma)$, which may be complex given the rule. This example is excluded by (S2) rather than by $\nu$: a random perturbation of the code destroys the match between code and machine, so no offspring is formed, and the code is not heritable.
\end{enumerate}
\end{remark}

\begin{remark}[Limitations of the amount $\nu$]\label{rem:search}
Unlike the conditions of Definition \ref{def:selfrep}, $\nu$ is a lower-bound quantity, and it has two limitations.

\emph{Unprovability.} By Chaitin's incompleteness theorem \cite{li2008}, a fixed consistent formal system can prove $K(z) > L$ for a specific string $z$ only for $L$ below a constant depending on the system. Since $\nu_k \ge L$ implies such a statement for every productive variant, lower bounds on $\nu$ for specific objects cannot be proved beyond a constant. Upper bounds on $\nu$ can be proved, and so can statements about most perturbations or most variants, by counting.

\emph{Search.} $\nu$ is measured by unbounded conditional complexity, so machinery that can be found by search in the rule counts as supplied by the rule. Suppose the productive variants for a given tape format could be identified by searching through candidate machines and testing them, e.g. as the first machine that, placed next to tapes of that format, reproduces. Then $\nu$ would be small even for the construction of von Neumann and Burks. Productivity is not decidable (it requires reproduction for every $N$), so it is not clear that such a short description exists; a finite search can be misled by machines that reproduce for a while and then fail. Distinguishing machinery that is explicitly written in the rule from machinery that is implicit in it but expensive to find would require resource bounds, e.g. Levin's time-bounded complexity or Bennett's logical depth \cite{li2008}. We have avoided resource bounds in the qualitative definition, but they may be unavoidable for an algorithmic measure of the amount of self-reproduction. The next subsection describes a measure that avoids both limitations without resource bounds.
\end{remark}

\subsection{Certified amount of self-reproduction}\label{sec:certified}

The difficulty with $\nu$ is that a short description of a machine may be a program that searches for it. We now measure instead the length of a \emph{proof} that the machine does its job. A finite search can confirm that a machine has produced $N$ copies by time $T$, but productivity requires copies for every $N$, and no finite computation establishes that. So a proof of productivity must contain a genuine argument, and a search cannot make it shorter.

Fix a consistent, recursively axiomatized theory $\mathcal{T}$ in which elementary arithmetic can be carried out, e.g. Peano arithmetic, in which finite objects (rules, configurations, boxes, arrays) are represented by numerals via a fixed coding. For a self-reproducer $p$ (given by $\phi$, $c_0$, $A$, $M$ and $k$) and a machine variant $m'$, let $\mathrm{Prod}_k(m')$ be the arithmetic sentence expressing that $m'$ is $k$-productive (Definition \ref{def:amount}). Since $K(z \,|\, w) \le k$ is a $\Sigma^0_1$ condition (some program of length at most $k$ outputs $z$ given $w$), and contents of boxes at time $t$ are computable from $c_0$ and $t$, $\mathrm{Prod}_k(m')$ is of the form $\forall N\, \exists t\, \exists (\text{boxes})\, \psi$ with $\psi$ in $\Sigma^0_1$, i.e. it is a $\Pi^0_2$ sentence. Let
\[
\Phi_p \;\equiv\; \exists m' \in \Sigma^M \; \mathrm{Prod}_k(m') \quad ,
\]
where $\phi$, $c_0$, $A$, $M$ and $k$ appear as numerals.

\begin{definition}[Certified amount of self-reproduction]\label{def:certified}
The certified amount of self-reproduction of $p$ relative to $\mathcal{T}$ is
\[
\nu^{\mathcal{T}}(p) \equiv \min \bigl\{ |\pi| \;\big|\; \pi \text{ is a } \mathcal{T}\text{-proof of } \Phi_p \bigr\} \quad ,
\]
the length of the shortest proof that some machinery can stand in for the parent's machine, with $\nu^{\mathcal{T}}(p) = \infty$ if there is no such proof.
\end{definition}

Under the propositions-as-types reading, with $\mathcal{T}$ a type theory, $\Phi_p$ is the dependent sum type $\sum_{m' : \Sigma^M} \mathrm{Prod}_k(m')$, and $\nu^{\mathcal{T}}(p)$ is the size of its smallest inhabitant: a machine together with a certificate that it works. The parent configuration, and hence the parent's machine, appears in the statement and is not charged for. What is charged for is the reason why some machine works.

\begin{proposition}[Basic properties]\label{prop:certified}\
\begin{enumerate}[label=(\roman*)]
\item For every $L$, the predicate $\nu^{\mathcal{T}}(p) \le L$ is decidable, uniformly in $p$ and $L$. In particular, $\nu^{\mathcal{T}}(p)$ is computable from $p$ whenever it is finite.
\item If $\mathcal{T}'$ extends $\mathcal{T}$ in the same language, then $\nu^{\mathcal{T}'}(p) \le \nu^{\mathcal{T}}(p)$.
\item If $\mathcal{T}$ is sound for arithmetic sentences (e.g. $\mathcal{T}$ is Peano arithmetic) and $\nu^{\mathcal{T}}(p) < \infty$, then some machine variant of $p$ is indeed $k$-productive.
\end{enumerate}
\end{proposition}
\begin{proof}
(i) There are finitely many strings of length at most $L$, and whether a string is a $\mathcal{T}$-proof of $\Phi_p$ is decidable, since $\mathcal{T}$ is recursively axiomatized. If $\nu^{\mathcal{T}}(p)$ is finite, checking all strings in order of length finds a shortest proof. (ii) A $\mathcal{T}$-proof is a $\mathcal{T}'$-proof. (iii) $\Phi_p$ is an arithmetic sentence, so this holds whenever $\mathcal{T}$ proves only arithmetic sentences that are true in the standard model; Peano arithmetic has this property.
\end{proof}

Part (i) is in sharp contrast with the algorithmic amount: by Chaitin's theorem, lower bounds on $\nu$ for specific objects cannot be proved beyond a constant, whereas lower bounds on $\nu^{\mathcal{T}}$ can be decided, at least in principle (the procedure is exhaustive and exponential in $L$). Part (iii) requires that $\mathcal{T}$ be sound for the sentences in question, not merely consistent.

\begin{remark}[Why search does not shorten certificates]\label{rem:reflection}
A finite search, e.g. a simulation of a candidate machine for $T$ steps, establishes a $\Sigma^0_1$ fact (``$N$ copies by time $T$''), and such facts can be proved in $\mathcal{T}$ by verifying the computation. The $\Pi^0_2$ statement $\mathrm{Prod}_k(m')$ does not follow from any finite collection of such facts, since a machine may produce copies for a while and then fail. A uniform way of turning successful searches into proofs would amount to a reflection principle for $\mathcal{T}$, namely $\mathcal{T}$-provability of $\mathrm{Prov}_{\mathcal{T}}(\ulcorner \Phi \urcorner) \to \Phi$; by Löb's theorem, $\mathcal{T}$ proves this for a sentence $\Phi$ only if it proves $\Phi$ itself. Thus search is blocked by a logical limitation, not by restricting computation, and in particular without resource bounds. This is the same mechanism by which typing blocks search when the objects searched for are proofs rather than data. Two caveats remain. First, the measure depends on $\mathcal{T}$: stronger theories can shorten proofs dramatically (by Gödel's speed-up phenomenon), so $\nu^{\mathcal{T}}$ is meaningful relative to a fixed, natural choice of $\mathcal{T}$. Second, in theories that check computations during proof checking (proofs by reflection, as in some type theories), decidable subgoals such as finite simulations can be discharged by short proofs; this does not affect the $\Pi^0_2$ content of $\mathrm{Prod}_k$, but it changes what is charged for.
\end{remark}

For the examples of Remark \ref{rem:catalyst}, the certified amount behaves like the algorithmic one. For a catalyst, for transcription by the rule, and for a key written into the rule, a short argument shows that a trivial machine works: the rule copies code next to it and recreates it next to every copy, and an induction on generations gives copies for every $N$. The length of this argument does not depend on the code, so $\nu^{\mathcal{T}}$ is small. For the construction of von Neumann and Burks, by contrast, a proof that some machine works must establish that a constructor builds what the tape describes and that a copier copies the tape, i.e. it must contain a correctness argument for the organs of the machine (Section \ref{sec:vonneumann}). We emphasize that these are expectations; to establish them requires the concrete rules.

\subsection{Description length, untyped and typed}\label{sec:mdl}

The two quantitative notions used in this section are instances of a single idea, minimum description length, applied in two different systems of descriptions.
\begin{itemize}
\item Conditional Kolmogorov complexity $K(y \,|\, x)$ is the length of the shortest description of $y$ in an untyped universal language, i.e. of a program that computes $y$ from $x$. It is the appropriate measure for determination between finite objects (Theorem \ref{thm:determination}), and it is used in this form in all the conditions of Definitions \ref{def:descent}, \ref{def:heritable} and \ref{def:selfrep}.
\item The certified amount $\nu^{\mathcal{T}}(p)$ is the length of the shortest description in a typed system: under the propositions-as-types reading, the size of the smallest inhabitant of the type $\Phi_p$. It is the appropriate measure for claims about infinitely many generations, which no finite computation establishes.
\end{itemize}
In an untyped universal language, a description may be a program that searches; this is harmless for upper bounds and harmful for lower bounds (Remark \ref{rem:search}). In a typed system, an inhabitant of a type expressing a $\Pi^0_2$ property cannot be produced by a finite search, by the absence of reflection (Remark \ref{rem:reflection}); this makes lower bounds meaningful, and even decidable (Proposition \ref{prop:certified}).

We can now state precisely the thesis that this section supports. \emph{Self-reproduction is to be defined, and its amount measured, by minimum description length relative to the dynamical law, in the system of descriptions appropriate to the claim, together with interventions.} The thesis has content because there are natural alternatives, which the results of this section show to be insufficient:
\begin{enumerate}[label=(\alph*)]
\item counts of offspring, or of non-quiescent cells: the parity rule proliferates although the copying is done by the rule (Proposition \ref{prop:fredkin});
\item rarity and other count-based quantities, such as functional information, the fraction of working machines, or the fraction of inherited perturbations: these attribute to the object information supplied by the rule (Proposition \ref{prop:keys});
\item description length that ignores the rule, such as the unconditional complexity of the machine: it makes the same misattribution (Proposition \ref{prop:keys});
\item any quantity computed from the actual run alone, including all algorithmic information quantities: it cannot distinguish a machine that is built from its description from one that is copied (Proposition \ref{prop:observational}), which is why interventions are needed.
\end{enumerate}
Within description length, Theorem \ref{thm:determination} shows that, for determination between finite objects, Kolmogorov complexity is forced up to logarithmic terms.

\begin{remark}[Invariance]\label{rem:invariance}
Kolmogorov complexity depends on the universal machine only up to an additive constant, and all conditions of Definition \ref{def:selfrep} inherit this invariance. Typed description length has no comparable invariance theorem: proof lengths in different theories may differ enormously, by Gödel's speed-up phenomenon, and $\nu^{\mathcal{T}}$ is invariant only for a fixed theory $\mathcal{T}$. Thus the qualitative definition is invariant in the strong sense of the Introduction, while the certified amount is meaningful relative to a fixed, natural choice of theory. Whether the distinction between small and large certified amounts is robust across a natural class of theories, e.g. theories whose proofs can be translated into each other with at most polynomial increase of length in the relevant cases, is an open question.
\end{remark}

\subsection{Why both information and interventions are needed}\label{sec:necessity}

We now show that the two ingredients of Definition \ref{def:selfrep}, information relative to the rule and interventions, cannot be dispensed with. Constructing the required examples as local rules of cellular automata would add nothing but length, so we work in an abstract model of reproduction by generations, in the spirit of the summary of the construction of von Neumann and Burks given in the Introduction.

\begin{definition}[Generational systems]\label{def:generational}
Let $\mathcal{T}$ and $\mathcal{M}$ be sets of binary strings (tapes and machines). An object is a pair $o = (\tau, m) \in \mathcal{T} \times \mathcal{M}$. A generational system is a total computable map $R$ assigning to each object a finite list $R(o)$ of objects, its offspring. The run of an object $o$ under $R$ is the rooted tree in which the root is labelled $o$ and the children of a node labelled $o'$ are labelled by the elements of $R(o')$. The record of the run is this labelled tree. An intervention replaces the root object $o$ by another object $o'$; the outcome of the intervention is the run of $o'$.
\end{definition}

The tape of an object is heritable if a perturbation of the tape of the root reappears in its children, and its machine is rebuilt from the tape (respectively copied) if, after a perturbation of the machine of the root, the children carry the original machine (respectively the perturbed one).

\begin{proposition}[Interventions are necessary]\label{prop:observational}
Let $\ell \ge 1$, $\mathcal{M} = \{0,1\}^\ell$, $\mathcal{T}$ the binary strings of length at least $\ell$, and $\mathrm{dec}(\tau)$ the first $\ell$ bits of $\tau$. Define
\[
R_1(\tau, m) = \bigl[ (\tau, \mathrm{dec}(\tau)), (\tau, \mathrm{dec}(\tau)) \bigr] \quad , \qquad R_2(\tau, m) = \bigl[ (\tau, m), (\tau, m) \bigr] \quad .
\]
For every $\tau$, the object $o = (\tau, \mathrm{dec}(\tau))$ has the same run under $R_1$ and under $R_2$: the full binary tree with all nodes labelled $o$. However, for every $m' \neq \mathrm{dec}(\tau)$, the children of $(\tau, m')$ are $(\tau, \mathrm{dec}(\tau))$ under $R_1$ and $(\tau, m')$ under $R_2$. Hence any criterion that depends only on the record of the run of $o$, including every algorithmic information quantity computed from the labels of the record, assigns the same value to $o$ under $R_1$ and under $R_2$. Yet under $R_1$ the machine is rebuilt from the tape, and under $R_2$ it is copied from the parent's machine.
\end{proposition}

\begin{proof}
Both maps send $o$ to $[o, o]$, so by induction both runs are the full binary tree labelled $o$. The statements about $(\tau, m')$ follow from the definitions. A criterion that depends only on the record takes the same value on equal records.
\end{proof}

The machine of $o$ is decodable from its tape in both systems, so it is informationally redundant. In the construction of von Neumann and Burks the same holds: the machine $D$ can be decoded from $d_D$. Whether such a machine is built or copied is a counterfactual property, which can be detected by interventions and not otherwise. This is a deterministic instance of the familiar fact that observed (conditional) independences underdetermine causal structure \cite{pearl2009}. In Section \ref{sec:ci} the same phenomenon appeared in a different form: by Proposition \ref{prop:collapse}, the weak dependence of a function at a point depends only on the value of the function there, not on how the function responds to changes of its argument.

\begin{proposition}[Information relative to the rule is necessary]\label{prop:keys}
Let $r$ be a fixed program which, given $w \in \{0,1\}^L$, computes the generational system
\[
R_w(\tau, m) = \begin{cases} \bigl[ (\tau, w), (\tau, w) \bigr] & \text{if } m = w, \\ [\;] & \text{otherwise,} \end{cases}
\]
and describe $R_w$ by the string $\langle r, w\rangle$. Let $o_w = (\tau, w)$ for a fixed tape $\tau$. Then:
\begin{enumerate}[label=(\roman*)]
\item the runs of $o_w$ under $R_w$, for different $w$, are isomorphic via the relabelling $w \mapsto w'$; in each, the root has two children and the run is infinite, the tape is heritable, perturbations of the machine are not inherited, and exactly a fraction $2^{-L}$ of the machine variants $m' \in \{0,1\}^L$ are productive;
\item for all but fewer than $2^{L - j}$ strings $w$, $K(w) \ge L - j$;
\item there is a constant $c$, independent of $L$ and $w$, such that $K(w \,|\, \langle r, w\rangle) \le c$: the only productive machine is determined by the rule.
\end{enumerate}
\end{proposition}

\begin{proof}
(i) The relabelling of $w$ by $w'$ maps $R_w$ to $R_{w'}$ and $o_w$ to $o_{w'}$. The root $(\tau, w)$ has the two children $(\tau, w)$, so the run is the full binary tree. Replacing $\tau$ by $\tau'$ gives children $(\tau', w)$, which carry $\tau'$. Replacing $w$ by $m' \neq w$ gives no children, so no information about $m'$ is inherited, and $m'$ is productive only if $m' = w$. (ii) There are fewer than $2^{L-j}$ programs of length less than $L - j$. (iii) $w$ is the second component of $\langle r, w \rangle$.
\end{proof}

By (i), any criterion that is invariant under relabelling, in particular any criterion based on counts of offspring, of productive variants, or of inherited perturbations, treats all the systems $R_w$ alike, and attributes to the object the $L$ bits of specificity reflected in the fraction $2^{-L}$. By (ii), so does the unconditional complexity of the machine, for most $w$. By (iii), all of these bits are supplied by the rule: conditioned on the rule, the simplest productive machine has complexity $O(1)$, which is the value of $\nu$. Only a measure of information relative to the rule makes this distinction.

Propositions \ref{prop:observational} and \ref{prop:keys} support the following conclusion. Algorithmic information alone, computed from the actual run, does not suffice to define self-reproduction, since construction and copying differ only counterfactually; and no criterion that ignores the information supplied by the rule suffices either. Definition \ref{def:selfrep} combines information relative to the rule with interventions. We do not claim that these two ingredients settle every question; in particular, Remark \ref{rem:search} shows that the quantitative measure $\nu$ inherits the limitations of unbounded algorithmic information.

\subsection{The construction of von Neumann and Burks}\label{sec:vonneumann}

The self-reproducing machine of von Neumann and Burks is the example that Definition \ref{def:selfrep} is meant to include. We do not prove that it does; here we isolate what must be true for it to do so. We first reduce Definition \ref{def:selfrep}, for objects that reproduce by making exact copies of themselves in an otherwise empty space, to explicit hypotheses.

\begin{proposition}[Exact replication]\label{prop:exact}
There is a constant $c'$, depending only on $U$ and the encodings, with the following property. Let $k \ge c'$, let $c_0$ be a finite configuration that vanishes outside a box $A$, so that $A$ is delimited in $c_0$, and let $A = T \sqcup M$ be a decomposition. Suppose that
\begin{enumerate}[label=(H\arabic*)]
\item \emph{(exact proliferation)} for every $N$ there are a time $t$ and $N$ pairwise disjoint boxes $A + s$, delimited in $c_t$, with $c_t|_{A+s}$ the translate of $c_0|_A$;
\item \emph{(heritability)} at one of these boxes, $T$ is $k$-heritable and $M$ is $k$-somatic;
\item \emph{(inert code)} condition (S4) holds.
\end{enumerate}
Then the parent $A$ is a $(c, k)$-self-reproducer with $c = K(m \,|\, \langle \phi, A, \gamma\rangle)$, the original machine is $c'$-productive, and $\nu_{c'} \le K(m \,|\, \langle \phi, A, M\rangle)$.
\end{proposition}

\begin{proof}
Let $B = A + s$ be one of the boxes in (H1), and $\theta = \langle \phi, t, A, B\rangle$. The array $v = c_t|_B$ is the array $x = c_0|_A$, and the frame determines $s$. Hence $v$ is computable from $\langle\theta, x\rangle$ and $x$ from $\langle\theta, v\rangle$ by a fixed program, which gives (D1) and (D3). Since $c_0$ vanishes outside $A$, the environment $y = c_0|_E$ is zero and computable from $\theta$, so $\kappa_\theta(c_0) \ge K(v \,|\, \theta) - O(1)$, which is (D2). Thus each such box is a $c'$-descendant, hence a $k$-descendant, and (H1) gives (S1). (S2) is (H2), (S3) holds by the choice of $c$, and (S4) is (H3). The same argument, applied to the unperturbed run, shows that $m$ is $c'$-productive.
\end{proof}

The quantity $c = K(m \,|\, \langle \phi, A, \gamma\rangle)$ is small when the machine can be decoded from its own description; it is then of the order of the complexity of the decoding procedure, independently of the size of the machine. For the construction of von Neumann and Burks, the following points must be settled.

\begin{enumerate}[label=(V\arabic*)]
\item \emph{A precise configuration.} Von Neumann's design was not specified to the last cell in the 29-state rule; complete implementations, e.g. that of Pesavento \cite{pesavento1995}, use extended rule sets. One must fix a rule and an initial configuration $c_0$ consisting of the machine $D$ together with its tape $d_D$, in an otherwise empty space.
\item \emph{Exact proliferation (H1).} The machine must complete the construction of an exact copy of itself and of its tape, the copy must in turn reproduce, and the offspring must be placed so that they never collide and are eventually delimited by quiescent cells. In two dimensions, the absence of collisions between generations is a genuine requirement on the design (e.g. on the direction of the construction arm), not an automatic consequence of reproduction. This is the main point to be proved, presumably by a modular argument about the organs of the machine rather than by simulation, since simulation can only verify finitely many generations.
\item \emph{Heritability (H2).} With $T$ the cells of the tape and $M$ the remaining cells of the box: a random change of the tape must reappear in the first offspring, whose tape is a copy of the parent's tape; and a random change of the machine must not, either because the damaged machine fails to reproduce or because the offspring's machine is rebuilt from the intact tape. Both are statements about most perturbations, and can be established by counting once the relevant organs are understood.
\item \emph{Inert code (H3).} A tape without the machine must not proliferate. Since the tape is a passive line of cells, this amounts to checking that tape cells are not changed by the rule in the absence of excitation, which we expect to be immediate.
\item \emph{Decodability.} Since $d_D$ is a description of $D$ in a fixed format, $D$ is computable from $d_D$ by a program of length about the complexity of that format, so $c$ is of that order by design.
\end{enumerate}

\begin{conjecture}\label{conj:vonneumann}
There is a complete implementation of the self-reproducing machine of von Neumann and Burks which satisfies (H1)--(H3). Consequently, by Proposition \ref{prop:exact}, it is a $(c,k)$-self-reproducer, with $c$ of the order of the complexity of its description format and $k$ a constant.
\end{conjecture}

The conjecture is qualitative. A proof of it in a theory $\mathcal{T}$, which by (V2) we expect to be a modular argument about the organs of the machine, yields at the same time a $\mathcal{T}$-proof of $\Phi_p$, hence an upper bound on the certified amount $\nu^{\mathcal{T}}$. We expect the certified amount to be of the order of the length of such a correctness argument, and in particular much larger than for the catalysts and keys of Remark \ref{rem:catalyst}; by Proposition \ref{prop:certified}(i), this is in principle decidable for any given bound. For the algorithmic amount $\nu$, we do not conjecture a lower bound: by Remark \ref{rem:search}, such a bound could not be proved beyond a constant, and it is not even clear that it holds. Only the upper bound $\nu \le K(m \,|\, \langle \phi, A, M\rangle)$ of Proposition \ref{prop:exact} is assured.

The constituents of $D$ are excluded as follows. The constructor alone, $(A, d_A) \to A$, does not copy its description: the offspring has no code, so it carries no copy of the parent and cannot be a descendant, and if the environment supplies descriptions, (D2) fails. The copier alone, $(B, d_X) \to d_X$, either contains no machine (if the object is $d_X$, which then fails (S4), as in the parity rule) or does not reproduce its machine (if the object is the pair $(B, d_X)$, the offspring $d_X$ cannot be a descendant, since the parent is not recoverable from it, so (D3) fails).

\section*{Declaration of generative AI use}

This manuscript was prepared with substantial assistance from Claude (Anthropic; model Claude Opus 5.5), used through a conversational interface in November 2026. The division of work was as follows.

The author posed the research question; introduced the notion of weak dependence (Definition \ref{def:weak} and the partial derivatives (\ref{eq:weak})), Theorem \ref{thm:one} and the argument of Fig. \ref{fig1}; wrote the initial draft of the Introduction and of Section 2; and directed the development of the project, including the proposals to separate code from machinery by typing, to replace non-informational conditions by information-based ones, and to measure complexity in typed settings by description length within the typed system.

The AI system, in response to the author's requests: observed that weak dependence reduces to algorithmic mutual information (Proposition \ref{prop:collapse}); formulated and proved Propositions \ref{prop:partial}--\ref{prop:busybeaver}, Lemma \ref{lem:key}, Theorems \ref{thm:effective}, \ref{thm:semigraphoid}, \ref{thm:intersection} and \ref{thm:determination}, and the material of Appendices \ref{app:further} and \ref{app:sorted}; proposed and wrote the definitions, propositions and proofs of Section 3; ran numerical checks of several information-theoretic statements (the influence identity, the intersection and Poincar\'e-type inequalities, and the rank scaling of Proposition \ref{prop:rank}); suggested references and compiled their bibliographic entries; and edited the language and structure of the whole manuscript, including the author's original text.

\appendix

\section{Proofs and further material for Section 2}\label{app:further}

\subsection{Proof of Theorem \ref{thm:semigraphoid}}\label{app:proofs}

\begin{proof}
Recall that $I(\cdot:\cdot\,|\,\cdot) \ge -c$ and is symmetric up to $c$ (by (F2)); this gives (C1$'$).
(C2$'$): since $\pair{x,u}$ and $x$ are computable from each other, $K(\pair{x,u}\,|\,z) \eqp K(x\,|\,z)$ and $K(\pair{\pair{x,u},y}\,|\,z) \eqp K(\pair{x,y}\,|\,z)$, hence $I(x:y\,|\,z) \eqp I(\pair{u,x} : y \,|\, z)$. By Lemma \ref{lem:chain} (with symmetry) $I(\pair{u,x}:y\,|\,z) \ge I(u:y\,|\,z) + I(x:y\,|\,\pair{z,u}) - \delta_n \ge I(u:y\,|\,z) - c - \delta_n$.
(C3$'$): as before, $I(x:y\,|\,z) \eqp I(x:\pair{u,y}\,|\,z) \ge I(x:u\,|\,z) + I(x:y\,|\,\pair{u,z}) - \delta_n \ge I(x:y\,|\,\pair{u,z}) - c - \delta_n$.
(C4$'$): by Lemma \ref{lem:chain}, $I(x:\pair{y,w}\,|\,z) \le I(x:y\,|\,z) + I(x:w\,|\,\pair{y,z}) + \delta_n \le k + l + \delta_n$, and $\pair{y,w}$, $\pair{w,y}$ are computable from each other.
In each case the additive constants are absorbed into $\delta_n$ by enlarging $c$.
\end{proof}

\subsection{Abstract formulation}

\begin{remark}[Abstract formulation]\label{rem:separoid}
In the language of Dawid \cite{dawid2001}, let $V$ be the set of strings with the preorder $u \le_k x \iff K(u\,|\,x) \le k$ and join $x \vee y = \pair{x,y}$; this is the semilattice studied in \cite{muchnik2002}. Theorems \ref{thm:semigraphoid} and \ref{thm:intersection} state that $\independent_k$ is a \emph{graded} separoid on $V$: each separoid axiom holds after the grade $k$ is increased by a quantity that is additive in the grades of the hypotheses and logarithmic in the lengths of the strings. (In (C2$'$) and (C3$'$), one may replace ``$u = h(x)$'' by ``$u \le_j x$'', at the cost of an additional $j$ in the grade, by the same proof.) These losses disappear in the asymptotic limit. For infinite binary sequences $\mathbf{x},\mathbf{y},\mathbf{z}$, define $\mathbf{x} \independent \mathbf{y} \,|\, \mathbf{z}$ if $I(\mathbf{x}_{1:n} : \mathbf{y}_{1:n} \,|\, \mathbf{z}_{1:n}) = o(n)$, and $\mathbf{u} \le \mathbf{x}$ if $K(\mathbf{u}_{1:n} \,|\, \mathbf{x}_{1:n}) = o(n)$. Then the same proofs show that (C1)--(C4) hold exactly, and (C5) holds exactly whenever $\mathbf{y} \independent \mathbf{z}$. More generally, linear information inequalities valid for Shannon entropy are valid for Kolmogorov complexity up to logarithmic terms, and conversely \cite{hammer2000}. This is why the probabilistic reasoning behind (C1)--(C5) transfers to the algorithmic setting.
\end{remark}

\subsection{The intersection property and common information}\label{app:intersection}

The proviso in (C5) and the argument of Fig. \ref{fig1} have an exact probabilistic counterpart. For finite-valued random variables, $X \independent Y \,|\, Z$ means $P(x \,|\, y,z) = P(x \,|\, z)$ whenever $P(y,z) > 0$. Let $G$ be the bipartite graph whose vertices are the values $y$ with $P(y)>0$ and the values $z$ with $P(z) > 0$, with an edge $\{y,z\}$ whenever $P(y,z) > 0$, and let $C(y,z)$ be the connected component of $G$ containing the edge $\{y,z\}$.

\begin{proposition}\label{prop:prob-intersection}
If $X \independent Y \,|\, Z$ and $X \independent Z \,|\, Y$, then $X \independent (Y,Z) \,|\, C(Y,Z)$. In particular, if $G$ is connected (e.g. if $P$ is strictly positive), then $X \independent (Y,Z)$.
\end{proposition}
\begin{proof}
For every edge $\{y,z\}$ of $G$ we have $P(x\,|\,z) = P(x\,|\,y,z) = P(x\,|\,y)$ for all $x$. Hence the function $x \mapsto P(x\,|\,\cdot)$ takes the same value at the two endpoints of every edge, and therefore at all vertices of a connected component. So $P(x\,|\,y,z) = \phi_{C(y,z)}(x)$ for some functions $\phi_C$. Since $C(Y,Z)$ is a function of $(Y,Z)$, this means $P(x \,|\, y,z) = P(x \,|\, C(y,z))$, i.e. $X \independent (Y,Z) \,|\, C(Y,Z)$. If $G$ is connected, $C$ is constant.
\end{proof}

\begin{remark}\label{rem:commoninfo}
The proof of Proposition \ref{prop:prob-intersection} is the same ``perpendicular lines cross'' argument as Lemma \ref{lem:good}: the conditional distribution is propagated along a path that alternately moves in the $y$- and in the $z$-direction. The same mechanism characterizes the intersection property for continuous densities in terms of path-connectedness of supports \cite{peters2015}, and for a function of two real variables with vanishing partial derivatives it gives constancy on each component of the domain that is connected by axis-parallel paths. The variable $C(Y,Z)$ is the Gács--Körner common information of $Y$ and $Z$ \cite{gacskorner1973}: functions $g(Y) = g'(Z)$ (almost surely) correspond exactly to functions of $C$. The algorithmic analogue is
\[
C_k(x,y) \equiv \max \{ K(w) \,|\, K(w\,|\,x) \le k, \; K(w\,|\,y) \le k \} \quad .
\]
By Proposition \ref{prop:partial}(ii), if $\partial^{\,|}_x f$ and $\partial^{\,|}_y f$ at $(x_0,y_0)$ are at most $k - c$ (for a suitable constant $c$), then $v$ itself is a candidate $w$, so $\partial_z f\,|_{\pair{x_0,y_0}} \lep C_k(x_0,y_0)$. This is the sharpest form of the proviso in (C5) for weak dependence. It can be much stronger than Theorem \ref{thm:effective}, because common information can be far smaller than mutual information \cite{gacskorner1973, muchnik2002}. In Dawid's terms, this is the additional axiom of a \emph{strong} separoid, which involves the meet $y \wedge z$ (here, the common information) in place of the join \cite{dawid2001}.
\end{remark}

\begin{question}
Does the strong-separoid form of the intersection property hold approximately for algorithmic conditional independence in general, i.e. is $I(x : \pair{y,z})$ bounded by $I(x:y\,|\,z) + I(x:z\,|\,y)$ plus a function of a suitable common information of $y$ and $z$ (rather than of $I(y:z)$), also when $x$ is not computable from $\pair{y,z}$?
\end{question}

\subsection{Analogy with differentiation}\label{sec:derivative}

We now make precise in which sense $f'(x)\,|_{x_0}$ and $\partial_x f$ behave like derivatives. The closest analogue is not the value of the derivative but its \emph{rank}: $f'(x)\,|_{x_0} \eqp I(x_0 : f(x_0))$ measures how many ``independent directions'' of variation of the argument (bits of information) reach the value. The dictionary is summarized in Table \ref{tab:dictionary}; we now justify its entries.

\begin{table}[h]
\centering
\small
\begin{tabular}{@{}p{0.47\textwidth}p{0.47\textwidth}@{}}
\toprule
Smooth map $f : \mathbb{R}^n \to \mathbb{R}^m$ & Computable $f$, algorithmic setting \\
\midrule
$\operatorname{rank} Df(x_0)$ & $f'(x)\,|_{x_0} \eqp I(x_0 : f(x_0)) \eqp K(f(x_0))$ \\
$\dim \ker Df(x_0)$ & $K(x_0 \,|\, f(x_0)^*)$ (information lost) \\
rank--nullity theorem & $K(x_0) \eqp f'(x)\,|_{x_0} + K(x_0 \,|\, f(x_0)^*)$ \\
$\operatorname{rank} D(g\circ f) \le \min(\operatorname{rank} Dg, \operatorname{rank} Df)$ & Theorem \ref{thm:one}(i), (C2) \\
$\operatorname{rank} \partial_x f(x_0,y_0)$ & $\partial^{\,|}_x f \,|_{x_0,y_0} \eqp I(x_0 : v \,|\, y_0)$ \\
$\operatorname{rank}[\partial_x f \,|\, \partial_y f] \le \operatorname{rank} \partial_x f + \operatorname{rank} \partial_y f$ & Theorem \ref{thm:effective}, (C5) \\
$\partial_x f = \partial_y f = 0 \Rightarrow f$ locally constant & Theorem \ref{thm:one}(ii), requires $x_0 \independent y_0$ \\
$\log|\det Df|$ is additive under composition & information loss is additive (Proposition \ref{prop:ranknullity}) \\
\bottomrule
\end{tabular}
\caption{Dictionary between differentiation and weak dependence.}
\label{tab:dictionary}
\end{table}

\begin{proposition}[Rank--nullity]\label{prop:ranknullity}
For computable $f$ and all $x_0$, $K(x_0) \eqp f'(x)\,|_{x_0} + K(x_0 \,|\, f(x_0)^*)$. Consequently the information loss $L(f, x_0) \equiv K(x_0) - K(f(x_0))$ satisfies $L(f,x_0) \eqp K(x_0 \,|\, f(x_0)^*) \gep 0$ and, for computable $g$, $L(g \circ f, x_0) = L(f, x_0) + L(g, f(x_0))$.
\end{proposition}
\begin{proof}
By (F3), $K(\pair{f(x_0), x_0}) \eqp K(f(x_0)) + K(x_0 \,|\, f(x_0)^*)$, and $K(\pair{f(x_0), x_0}) \eqp K(x_0)$ by (F2). Now use Proposition \ref{prop:collapse}. The additivity of $L$ is immediate from its definition (the sum telescopes).
\end{proof}

The additivity of information loss under composition is the counterpart of the chain rule $\log |\det D(g\circ f)| = \log|\det Dg \circ f| + \log |\det Df|$. Indeed, if $X$ has a density and $f$ is a diffeomorphism, the change of differential entropy is $h(f(X)) - h(X) = \mathbb{E} \log |\det Df(X)|$ \cite{cover2006}. In probability theory, Shannon entropy is characterized, up to a constant factor, as the only information loss that is functorial (additive under composition), convex-linear and continuous \cite{baez2011}. From a homological point of view, the chain rule makes entropy a $1$-cocycle, a Leibniz-type identity \cite{baudot2015}. The analogy with derivations persists even modulo a prime \cite{leinster2019}.

The rank interpretation can be made quantitative for smooth maps. For $\varepsilon > 0$ and $w \in \mathbb{R}^m$ let $[w]_\varepsilon \equiv (\lfloor w_1/\varepsilon \rfloor, \ldots, \lfloor w_m/\varepsilon \rfloor)$ be the quantization of $w$ at scale $\varepsilon$, and let $H$ denote Shannon entropy in bits.

\begin{proposition}[Information flow recovers the rank]\label{prop:rank}
Let $f : \mathbb{R}^n \to \mathbb{R}^m$ be twice continuously differentiable in a neighbourhood of $x_0$, and $r = \operatorname{rank} Df(x_0)$. For $\delta > 0$ let $X$ be uniformly distributed on the cube $x_0 + [-\delta/2, \delta/2]^n$, and $\varepsilon = \delta^{3/2}$. Then there is a constant $C$ such that, for all sufficiently small $\delta$,
\[
\Bigl| \, I\bigl(X ; [f(X)]_\varepsilon\bigr) - \frac{r}{2} \log_2 \frac{1}{\delta} \, \Bigr| \le C \quad .
\]
\end{proposition}
\begin{proof}
Since $[f(X)]_\varepsilon$ is a function of $X$, $I(X;[f(X)]_\varepsilon) = H([f(X)]_\varepsilon)$. Let $A = Df(x_0)$, $u = X - x_0$, so $|u| \le \sqrt{n}\,\delta/2$. Let $M$ bound the second derivatives of $f$ near $x_0$. By Taylor's theorem, $f(X) = f(x_0) + Au + e$ with $|e| \le M n \delta^2 / 8$, and for $\delta$ small enough $M n \delta^2/8 \le \delta^{3/2} = \varepsilon$. If $r = 0$, then $f(X)$ lies within $\varepsilon$ of $f(x_0)$, so $[f(X)]_\varepsilon$ takes at most $3^m$ values, and the claim holds. Assume $r \ge 1$, and let $\sigma_1 \ge \ldots \ge \sigma_r > 0$ be the non-zero singular values of $A$.

\emph{Upper bound.} $Au$ lies in the $r$-dimensional ball $\operatorname{im}A \cap B(0, \rho)$ with $\rho = \sigma_1 \sqrt{n}\,\delta/2$. This ball is covered by at most $(1 + 2\rho/\varepsilon)^r$ balls of radius $\varepsilon$, so $f(X)$ lies in the union of at most $(1 + 2\rho/\varepsilon)^r$ balls of radius $2\varepsilon$. Each such ball meets at most $5^m$ quantization cells. Hence $[f(X)]_\varepsilon$ takes at most $5^m (1 + \sigma_1\sqrt{n}\,\delta^{-1/2})^r \le C_1 \delta^{-r/2}$ values, and $H \le \frac{r}{2}\log_2\frac1\delta + \log_2 C_1$.

\emph{Lower bound.} Fix a quantization cell $Q$, of diameter $\sqrt{m}\,\varepsilon$. If $f(X) \in Q$, then $Au$ lies in a ball of radius $s_0 = (\sqrt{m}+1)\varepsilon$. Write $u = u_\parallel + u_\perp$ with $u_\parallel \in W \equiv (\ker A)^\perp$, $\dim W = r$. Since $Au = Au_\parallel$ and $|Aw| \ge \sigma_r |w|$ for $w \in W$, any two admissible values of $u_\parallel$ are at distance at most $2s_0/\sigma_r$, so $u_\parallel$ lies in a ball of radius $s = 2 s_0/\sigma_r$ in $W$. The orthogonal projection of $X - x_0$ onto $W$ has a density bounded by $\omega_{n-r} (\sqrt{n}\,\delta)^{n-r} / \delta^n$, where $\omega_j$ is the volume of the unit ball in $\mathbb{R}^j$, because each fibre of the projection is a convex subset of the cube, of diameter at most $\sqrt{n}\,\delta$. Hence
\[
P(f(X) \in Q) \le \omega_r s^r \cdot \omega_{n-r} n^{(n-r)/2} \delta^{-r} = C_2 (\varepsilon/\delta)^r = C_2 \delta^{r/2} \quad ,
\]
with $C_2$ independent of $\delta$ and $Q$. Therefore $H \ge -\log_2 \max_Q P(f(X) \in Q) \ge \frac{r}{2}\log_2\frac1\delta - \log_2 C_2$.
\end{proof}

Thus, at small scales, the mutual information between a perturbation of the argument and the (finely observed) perturbation of the value grows like $\operatorname{rank} Df(x_0)$ times $\frac12\log_2(1/\delta)$; heuristically, the bounded remainder is governed by $\sum_{i \le r} \log_2 \sigma_i$, the logarithm of the pseudo-determinant of $Df(x_0)$ (we do not use this). Accordingly, (C2) for mutual information (data processing) corresponds to the rank inequality for compositions. With $X$ and $Y$ independent and uniform on small cubes, (C5) corresponds to $\operatorname{rank}[\partial_x f \,|\, \partial_y f] \le \operatorname{rank}\partial_x f + \operatorname{rank}\partial_y f$. If instead the perturbations of $x$ and $y$ are dependent (e.g. confined to the diagonal $x = y$), the partial derivatives no longer control the total variation, which is the analogue of the role of $I(x_0:y_0)$ in Theorem \ref{thm:effective}.

On discrete domains, conditional mutual information coincides with the classical notion of influence of a variable, i.e. with the probability that a discrete partial derivative is non-zero \cite{odonnell2014}. For $f : \{0,1\}^n \to \{0,1\}$ let $\partial_i f(x) \equiv f(x^{i \leftarrow 1}) - f(x^{i\leftarrow 0})$, where $x^{i\leftarrow b}$ is $x$ with $i$-th coordinate set to $b$, and $\mathrm{Inf}_i(f) \equiv P(\partial_i f(X) \neq 0)$ for $X$ uniform on $\{0,1\}^n$. Write $X_{-i}$ for $X$ with the $i$-th coordinate removed.

\begin{proposition}[Influence]\label{prop:influence}
For $X$ uniform on $\{0,1\}^n$, $I(X_i ; f(X) \,|\, X_{-i}) = \mathrm{Inf}_i(f)$ (in bits).
\end{proposition}
\begin{proof}
$I(X_i; f(X)\,|\,X_{-i}) = H(f(X)\,|\,X_{-i}) - H(f(X)\,|\,X) = H(f(X)\,|\,X_{-i})$. Condition on $X_{-i} = x_{-i}$; the event $\partial_i f \ne 0$ depends only on $x_{-i}$. If $\partial_i f(x) = 0$, then $f(X)$ is determined by $x_{-i}$ and the conditional entropy is $0$. Otherwise $f(X)$ is a bijective function of $X_i$, which is uniform and independent of $X_{-i}$, so the conditional entropy is $1$ bit. Averaging over $x_{-i}$ gives $P(\partial_i f(X) \neq 0)$.
\end{proof}

Theorem \ref{thm:effective} has a probabilistic counterpart for any number of variables. It is analogous to the Poincaré (Efron--Stein) inequality, which bounds the variance of a function by the sum of the influences of its variables \cite{odonnell2014}. For a finite-valued random vector $X = (X_1, \ldots, X_n)$ define the dual total correlation $D(X) \equiv H(X) - \sum_i H(X_i \,|\, X_{-i}) \ge 0$; for $n = 2$, $D(X) = I(X_1 ; X_2)$, and $D(X) = 0$ if the $X_i$ are independent.

\begin{proposition}[Poincaré-type inequality]\label{prop:poincare}
For any function $F = f(X)$,
\[
I(F ; X) \le \sum_{i=1}^n I(X_i ; F \,|\, X_{-i}) + D(X) \quad .
\]
In particular, for $X$ uniform on $\{0,1\}^n$ and $f$ Boolean, $H(f(X)) \le \sum_i \mathrm{Inf}_i(f)$.
\end{proposition}
\begin{proof}
Write $X_{<i} = (X_1, \ldots, X_{i-1})$ and $X_{>i} = (X_{i+1}, \ldots, X_n)$. By the chain rule, $I(F;X) = \sum_i I(X_i ; F \,|\, X_{<i})$, and
\[
I(X_i ; F \,|\, X_{<i}) \le I(X_i ; (F, X_{>i}) \,|\, X_{<i}) = I(X_i ; X_{>i} \,|\, X_{<i}) + I(X_i ; F \,|\, X_{-i}) \quad .
\]
Finally, $\sum_i I(X_i ; X_{>i} \,|\, X_{<i}) = \sum_i [H(X_i\,|\,X_{<i}) - H(X_i \,|\, X_{-i})] = H(X) - \sum_i H(X_i\,|\,X_{-i}) = D(X)$. The last statement follows from Proposition \ref{prop:influence}, since $I(F;X) = H(F)$ and $D(X) = 0$.
\end{proof}

For $n = 2$ and $F = f(X_1, X_2)$, Proposition \ref{prop:poincare} reads $I(F ; (X_1,X_2)) \le I(X_1; F \,|\, X_2) + I(X_2 ; F \,|\, X_1) + I(X_1;X_2)$, which is the probabilistic counterpart of Lemma \ref{lem:key}. The dependence term $D(X)$ cannot be omitted: if $X_1 = X_2$ is a uniform bit and $F = X_1$, the left-hand side is $1$ and the sum of the conditional terms is $0$. This is the probabilistic counterpart of Remark \ref{rem:condfails}.

\section{An alternative formulation: sorted automata}\label{app:sorted}

In Section \ref{sec:heritability}, code and machine are identified by interventions. Alternatively, the distinction can be made from the outset, by assigning to each state a sort. We record this formulation, which leads to a no-cloning condition expressed as an algorithmic conditional independence, and to a proof that, under this condition, machines present in two copies are determined by code.

\begin{definition}[Sorted automaton]\label{def:sorted}
A sorting of the alphabet is a partition $\Sigma = \{0\} \sqcup \Sigma_C \sqcup \Sigma_M$ into the quiescent state, code states and machine states. A sorted automaton is a cellular automaton together with a sorting; we write $\hat\phi = \langle \phi, \Sigma_C, \Sigma_M\rangle$ for its sorted rule. For a configuration $c$ and a box $B$, let $c^C|_B$ (respectively $c^M|_B$) be the array on $B$ that coincides with $c$ at cells whose state is a code (respectively machine) state, and is $0$ elsewhere. For a run $(c_t)$ we write
\[
C_t(B) = c_t^C|_B \quad , \qquad M_t(B) = c_t^M|_B \quad ,
\]
and $C_0 = c_0^C$ for the code part of the whole initial configuration (a finite configuration).
\end{definition}

The informal content of the distinction is the following discipline: information may be duplicated only in the form of code. Two pieces of machinery may share information only through code, never because one has been copied from the other. In algorithmic terms this is a conditional independence statement.

\begin{definition}[No-cloning, well-typed runs]\label{def:nocloning}
Let $c \ge 0$. The run of a finite configuration $c_0$ in a sorted automaton is $c$-well-typed if for every $t \ge 0$ and every pair of disjoint boxes $B_1$, $B_2$,
\[
I\bigl( M_t(B_1) : M_t(B_2) \,\big|\, \langle \Theta, C_0 \rangle \bigr) \le c \quad , \qquad \Theta = \langle \hat\phi, t, B_1, B_2 \rangle \quad ,
\]
i.e. $M_t(B_1) \independent_c M_t(B_2) \,|\, \langle \Theta, C_0\rangle$ in the notation of Definition \ref{def:ci}.
\end{definition}

The condition says that the code present at time $0$ (together with the sorted rule) is, up to $c$ bits, the only common source of information of any two pieces of machinery present at a later time. It is an upper bound on information. The constant $c$ must be at least of the order of the complexity of the ``genetic code'', i.e. of the procedure by which machines are decoded from descriptions; it should be compared with the complexities of the objects under consideration.

The following lemma shows that well-typedness follows from a static condition on the initial configuration: it suffices that all machinery present at time $0$ can be decoded from the code present at time $0$.

\begin{lemma}[A static sufficient condition]\label{lem:static}
There is a constant $c'$, depending only on $U$ and the encodings, such that for every sorted automaton and every finite configuration $c_0$, the run of $c_0$ is $c$-well-typed with
\[
c = K\bigl( c_0^M \,\big|\, \langle \hat\phi, C_0 \rangle \bigr) + c' \quad ,
\]
where $c_0^M$ is the machine part of the whole configuration $c_0$ (the finite configuration that coincides with $c_0$ at machine cells and is $0$ elsewhere).
\end{lemma}

\begin{proof}
Let $t$, $B_1$, $B_2$ be arbitrary, $\Theta = \langle \hat\phi, t, B_1, B_2\rangle$ and $Z = \langle \Theta, C_0\rangle$. For any strings, $I(x : y \,|\, z) \le K(x \,|\, z) + O(1)$, since $K(\langle x,y\rangle \,|\, z) \ge K(y \,|\, z) - O(1)$ by (F2). The configuration $c_0$ is computable from $C_0$ and $c_0^M$ (superpose them), and $M_t(B_1)$ is computable from $c_0$ and $\Theta$ (run the automaton for $t$ steps and restrict). Hence $K(M_t(B_1) \,|\, Z) \lep K(c_0^M \,|\, Z) \lep K(c_0^M \,|\, \langle \hat\phi, C_0\rangle)$, where the last step uses that $\hat\phi$ is computable from $\Theta$. Combining, $I(M_t(B_1) : M_t(B_2) \,|\, Z) \le K(c_0^M \,|\, \langle \hat\phi, C_0\rangle) + c'$.
\end{proof}

Conversely, by Proposition \ref{prop:nocloning} below, in a well-typed run every machine that is later present in two copies is determined by the code at time $0$. Definition \ref{def:nocloning} is nonetheless weaker than the static condition of Lemma \ref{lem:static}, since it allows the environment to contain machinery that is unrelated to any code, provided that pieces of it do not share information with each other.

\begin{proposition}[No-cloning forces construction from code]\label{prop:nocloning}
Suppose that the run of $c_0$ is $c$-well-typed. Let $A$ be a box, $m = M_0(A)$, and suppose that at some time $t$ there are disjoint boxes $B_1 = A + s_1$ and $B_2 = A + s_2$ such that $M_t(B_1)$ and $M_t(B_2)$ are both translates of $m$. Then
\[
K\bigl( m \,\big|\, \langle \hat\phi, A, C_0 \rangle \bigr) \le c + K\bigl( \langle t, s_1, s_2 \rangle \,\big|\, \langle \hat\phi, A \rangle \bigr) + c' \quad ,
\]
where $c'$ depends only on $U$ and the encodings.
\end{proposition}

\begin{proof}
Let $\Theta = \langle \hat\phi, t, B_1, B_2\rangle$ and $m_l = M_t(B_l)$. Given $\Theta$, each of $m_1$, $m_2$ is computable from the other (translate by $s_2 - s_1$, which is computable from $B_1$, $B_2$), and $m$ is computable from $m_1$ (translate back to $A = B_1 - s_1$, where $A$ and $s_1$ are computable from $B_1$ and the size of $A$, i.e. from $\Theta$ together with $\langle \hat \phi, A\rangle$). Hence, writing $Z = \langle \Theta, C_0\rangle$,
\[
I(m_1 : m_2 \,|\, Z) = K(m_1\,|\,Z) + K(m_2\,|\,Z) - K(\langle m_1, m_2\rangle\,|\,Z) \eqp K(m_1 \,|\, Z) \quad ,
\]
since $K(\langle m_1, m_2\rangle\,|\,Z) \eqp K(m_1\,|\,Z) \eqp K(m_2\,|\,Z)$ by (F2). By well-typedness, $K(m_1 \,|\, Z) \le c + O(1)$. Now $m$ is computable from $m_1$ given $Z$, so $K(m \,|\, Z) \le c + O(1)$. Finally, $Z$ is computable from $\langle \hat\phi, A, C_0\rangle$ together with $\langle t, s_1, s_2\rangle$, so by (F2)
\[
K(m \,|\, \langle \hat\phi, A, C_0\rangle) \lep K(m \,|\, Z) + K(\langle t, s_1, s_2 \rangle \,|\, \langle \hat\phi, A\rangle) \quad . \qedhere
\]
\end{proof}

Thus, in a well-typed run, a machine that is present in two copies is determined by the code (and the rule), up to $c$ bits and the information needed to locate the copies. The constructor $A$ of the Introduction is therefore not an additional assumption: it is forced by the typing. The term $K(\langle t, s_1, s_2\rangle \,|\, \langle \hat\phi, A\rangle)$ is typically logarithmic in $t$ and in the distances involved.

\bibliographystyle{unsrt}
\bibliography{kolm_additions}

\end{document}